\documentclass[acmsmall]{acmart}

\usepackage{enumerate, enumitem, xcolor}

\AtBeginDocument{%
  }

\copyrightyear{2024}
\acmYear{2024}
\setcopyright{rightsretained}
\acmConference[ICTD 2024]{The 13th International Conference on Information \& Communication Technologies and Development}{December 09--11, 2024}{Nairobi, Kenya}
\acmBooktitle{The 13th International Conference on Information \& Communication Technologies and Development (ICTD 2024), December 09--11, 2024, Nairobi, Kenya}
\acmPrice{}
\acmDOI{10.1145/3700794.3700802}
\acmISBN{979-8-4007-1041-4/24/12}

\begin{document}

\title[A Civics-oriented Approach to Understanding Intersectionally Marginalized Users' Experience with Hate Speech]{A Civics-oriented Approach to Understanding Intersectionally Marginalized Users' Experience with Hate Speech Online}


\author{Achhiya Sultana}
\authornote{Both authors contributed equally, and the student author is put first.}
\affiliation{%
  \department{Department of Computer Science \& Engineering}
  \institution{Independent University Bangladesh}
  \city{Dhaka}
  \country{Bangladesh}
}
\email{achhiyasets@iub.edu.bd}

\author{Dipto Das}
\authornotemark[1]
\affiliation{%
  \department{Department of Computer Science}
  \institution{University of Toronto}
  \city{Toronto}
  \state{Ontario}
  \country{Canada}
}
\additionalaffiliation{
  \department{Department of Information Science, }
  \institution{University of Colorado Boulder}
  \city{Boulder}
  \state{Colorado}
  \country{United States}
}
\email{diptodas@cs.toronto.edu}

\author{Saadia Binte Alam}
\affiliation{%
  \department{Department of Computer Science \& Engineering}
  \institution{Independent University Bangladesh}
  \city{Dhaka}
  \country{Bangladesh}
}
\email{saadiabinte@iub.edu.bd}

\author{Mohammad Shidujaman}
\affiliation{%
  \department{Department of Computer Science \& Engineering}
  \institution{Independent University Bangladesh}
  \city{Dhaka}
  \country{Bangladesh}
}
\email{shidujaman@iub.edu.bd}

\author{Syed Ishtiaque Ahmed}
\affiliation{%
  \department{Department of Computer Science}
  \institution{University of Toronto}
  \city{Toronto}
  \state{Ontario}
  \country{Canada}
}
\email{ishtiaque@cs.toronto.edu}

\renewcommand{\shortauthors}{}

\begin{abstract}
  While content moderation in online platforms marginalizes users in the Global South at large, users of certain identities are further marginalized. Such users often come from Indigenous ethnic minority groups or identify as women. Through a qualitative study based on 18 semi-structured interviews, this paper explores how such users' experiences with hate speech online in Bangladesh are shaped by their intersectional identities. Through a civics-oriented approach, we examined the spectrum of their legal status, membership, rights, and participation as users of online platforms. Drawing analogies with the concept of citizenship, we develop the concept of usership that offers a user-centered metaphor in studying moderation and platform governance.
\end{abstract}




\maketitle

\section{Introduction}
Online platforms face significant challenges related to platform governance mechanisms like content moderation, especially concerning hate speech. While this issue is pervasive worldwide, the perception of and attention to its severity varies across geopolitical locations, economic interests, and cultures~\cite{jiang2021understanding, srnicek2017platform}. Content moderation decisions in the Global South often lack the necessary promptness and contextual understanding~\cite{das2021jol, shahid2023decolonizing}. Incidents from time to time highlighted how such a lack of nuances results in a disproportionate restriction of users' freedom of expression, exacerbating distrust and reinforcing biased perceptions~\cite{bhatia2024context}. As people engage in sociopolitical discourses and access various opportunities through online platforms, addressing these disparities is crucial for fostering inclusive online environments. Instead of viewing users through a dichotomy of being included or excluded, such as respectively the ones from the Global North and the Global South, we adopt a civics-oriented framework to understand the experiences of intersectionally marginalized users (e.g., Indigenous ethnic minority communities) in the Global South.

The difficulty in addressing this problem lies in platforms' conventional approach of trying to conceptualize their "average user." Cross-cultural computing research in information and communication technology for development (ICTD), human-computer interaction (HCI), and other adjacent areas often use countries as a proxy for culture, shaping the expectation of the average users in that context~\cite{das2024identity}. Let's consider design decisions for an online platform in Bangladesh. In a country where 98\% people speak Bengali but with many accents and dialects~\cite{das2023toward}, and where societal norms are shaped by patriarchy~\cite{sultana2018design}, content moderation (e.g., identifying hate speech) would prioritize the perspective of urban Bangladeshi men from the Bengali ethnic group over that of non-Bengali ethnic minorities, women, and people who speak Bengali in non-normative ways. Through a postcolonial perspective, the latter group is identified as the subaltern--the further marginalized group within the colonially marginalized Global South. As computing platforms exhibit a certain colonial impulse~\cite{dourish2012ubicomp, irani2010postcolonial}, even the ones aiming to create inclusive spaces for non-English speaking users in the Global South, limit users' juridical and cultural inclusion, entitlements, and responsibilities.

Our approach builds on the concept of citizenship and its politics~\cite{stokke2017politics}, translating it into the space of ICTD through drawing analogies. Prior social computing research found the use of metaphors and analogies effective in understanding online platforms' affordances~\cite{bruckman2006new, semaan2015designing} and governance approaches~\cite{seering2022metaphors, das2021jol}. Following that line of social computing research where online platforms are often compared to public spheres~\cite{semaan2015designing, semaan2015navigating} or content moderators are structured as juries~\cite{fan2020digital}, we extend the analogy by describing users as citizens of these digital spaces. By drawing analogies with different components of citizenship in civics, we overcome the limitation of treating users homogeneously. Complementing prior ICTD studies on the marginalized users in the Global South~\cite{das2022understanding, sultana2022toleration, nova2019online} and building on an empirical study with subaltern users like Indigenous ethnic minorities (Adivasi) and women in Bangladesh and their encounters with hate speech, we propose a framework that provides a structured way to examine their experiences on those platforms.

We conducted semi-structured interviews with users from Indigenous ethnic minority communities and female users in Bangladesh. Our empirical research to understand their experience of interactions with other users from majority communities and platform policies revealed inconsistencies, societal biases, and challenges in how hate speech is collectively and algorithmically moderated, highlighting the need for reconsidering governance mechanisms and policies. We looked into the intersectionally marginalized subaltern users' perspectives through four interrelated components: (1) legal status determined by their juridical inclusion; (2) membership, which shapes and is shaped by their cultural inclusion; (3) rights as their entitlement to voice opinions; and (4) participation defined by their responsibilities on the platforms. While ICTD and social computing literature have previously studied the marginalization of women and minorities online, our study contributes to that scholarship by focusing on Bangladeshi female users and Indigenous communities.

In the discussion section, where we develop the concept of usership, we reflect on how this framework can effectively be translated to different online platforms and communities to center users' experiences. Therefore, our paper not only empirically contributes by foregrounding the multi-dimensional marginalization of Indigenous ethnic minorities and women in the Global South but also makes a theoretical contribution to studying governance on online platforms. The paper is structured as follows: we first review related work on intersectional identity-based marginalization, the power structure of governance, and content moderation (Section~\ref{sec:literature_review}) and detail our methodology, positionality and limitation (Section~\ref{sec:methods}). Then, we present the findings from the interviews while drawing analogies with components of citizenship (Section~\ref{sec:results}). In Section~\ref{sec:discussion}, we develop the usership framework and discuss its limitations and implications for ICTD and social computing research.
\section{Literature Review}\label{sec:literature_review}
This section discusses how people's ethnic and gender identities shape their intersectional marginalization and how governance gives a civic structure to such social relations of inclusion, exclusion, and power. Here, we contextualize our discussion with examples, entities, and events specific to Bangladesh and its demography. Later in this section, we explain how the notion of governance is studied in the context of user interaction and content moderation in online platforms.

\subsection{Subaltern as Intersectionally Marginalized Identities}

Identity--how we see ourselves and want others to see us as social and physical beings~\cite{erikson1968identity, gecas1982self, goffman1959presentation} mediates our interaction and relationship with others. Though it is often construed as an individuated concept, various social identities emerge centered around people's perceived membership in different groups defined across race, ethnicity, gender, nationality, etc.~\cite{tajfel1974social}. People are often marginalized or pushed to the peripheries of societies based on various dimensions of their identities. A mechanism through which the marginalization of various identities has been historically imposed and normalized is colonialism~\cite{das2023studying}--the process by which external forces settle, control, alter, and exploit local identities, resources, and cultural, social, political, and economic structures~\cite{loomba2002colonialism}. In colonized communities, it created identity hierarchies like racism, where certain races were privileged over the others~\cite{fanon2008black, fanon2007wretched}. Its pervasive impacts have transgenerationally shaped ``traditional" gender roles~\cite{sinha2017colonial} and normalized Western beauty standards by stigmatizing larger bodies~\cite{farrell2011fat} and darker skin tones~\cite{hunter2007persistent}. Thus, some communities are marginalized in multiple ways.

In questioning how colonization further and disproportionately marginalized certain groups within the colonized local communities, Spivak uses the concept of ``subaltern"~\cite{spivak2003can}. She described how both colonial and patriarchal structures silenced Bengali women and barred them from representing themselves within dominant discourses. Her work shows how viewing through the dichotomy of colonizer-colonized alone misses the full extent of women's marginalization in postcolonial societies. Similarly, in exploring how race and gender interact to create specific experiences of discrimination for Black women, which cannot be understood by looking at race and gender separately, Crenshaw used the lens of ``intersectionality", which argues that various social categories compound to create unique experiences of discrimination and privilege~\cite{crenshaw2013mapping, crenshaw2013demarginalizing}.

The absence of such an intersectional approach to understanding colonial marginalization overlooks and trivializes the experiences of the subalterns--the marginalized within the marginalized. For example, unlike the Americas, where Indigenous populations are often recognized based on historical continuity with pre-colonial societies~\cite{eubanks2018we}, scholars examined who qualifies as Indigenous, considering how Asia's diverse cultural and historical landscapes and contemporary emancipatory politics complicated its definitions~\cite{baviskar2013politics}. While in studying the colonial marginalization of larger local ethnolinguistic communities (e.g., Bengalis) in South Asia, researchers also recognized the intersectional oppression of smaller ethnic tribes and Adivasi groups (e.g., Chakma, Marma)~\cite{xaxa1999tribes, das2022collaborative}, the latter has received little attention in HCI, CSCW, and ICTD literature. Similarly, researchers in these spaces~\cite{hampton2021black, ghosh2023person, dosono2020decolonizing, narr2022coloniality} only nominally looked at the experiences of women of color and Indigenous backgrounds (e.g., stereotypes and fetishization) with and through technologies.

Such under-representation of certain identities in scientific research and scholarship is structurally intertwined with how these identities are marginalized or prioritized and their depictions are institutionalized by and within dominant power structures~\cite{ali2016brief, irani2010postcolonial, dourish2012ubicomp, said2014orientalism}, such as governance.

\subsection{Governance as Civic Structure of Inclusion and Power}
Governance refers to the systematic and structured processes through which authority and decisions are exercised, and civic affairs are managed, involving both governmental institutions and non-state administrative actors. In postcolonial nation-states, governance often reanimates colonial hierarchies, values, and authority~\cite{fanon2007wretched, ramnath2012decolonizing}, which scholars have critically examined from various angles. In the context of Bengal, Banerjee explained how British colonial historiography construed ethnic tribes or Adivasis as ``primitive" and outside the realm of modern civilization~\cite{banerjee1998politics}. Institutionalizing such perspectives, particularly those related to development and conservation, postcolonial governance policies historically marginalized ethnic minorities in India~\cite{guha2000unquiet}. Related to Tuck and Yang's concerns around Indigenous lands and epistemologies~\cite{tuck2012decolonization}, Guha's work highlights the adverse effects of state policies and corporate encroachments on Adivasi communities in India. Scholars documented the conflicts between Indigenous ways of life and modern development projects, how those accompanied by militarization led to Adivasis' displacement, loss of livelihoods, and sexual violence against Adivasi women, and grassroots Adivasi movements and resistances~\cite{guha2000unquiet, sundar2016burning}. In doing so, they advocated for inclusive and equitable approaches toward Indigenous rights.

The Adivasis in the Chittagong Hill Tracts (CHT) of Bangladesh have faced systematic state-sponsored violence and exploitation since before Bangladesh's emergence as a nation-state and throughout the periods of colonial subjugation. British colonizers encroached upon their lands for tea plantations and other economic ventures~\cite{rasul2007political}. Under Pakistani rule, the construction of the Kaptai Dam displaced thousands, seizing their fertile lands~\cite{zaman1982crisis}. Post-independence, the Bangladeshi government continued this legacy through ethnocide, militarization, and settlement programs~\cite{chakma2010post, hill2022muscular}. While the CHT Peace Accord aimed to address these issues by promising autonomy and land rights, its implementation has been inconsistent, leaving many underlying tensions unresolved~\cite{chakma2008assessing}. Especially, ``an upward trend" of and ``a culture of impunity" for sexual abuse, abduction, physical assault, and attempted rape of Indigenous women are matters of critical concern~\cite{iwgia2021violence}.

Decolonial scholars have examined how the assimilation of ethnic minorities in the paradigm of nation-states and citizenship can be problematic~\cite{ramnath2012decolonizing}. For example, despite being Bangladeshi citizens, Adivasis in the country were historically pressured to identify as ``Bengalis" instead of their ethnic identities like Chakma and Marma and are derogatively called \textit{Upajati}\footnote{The word translates to ``sub-nation," which resembles Frantz Fanon's quote about the Black people's cultural assimilation: ``To be Black is to be subhuman"~\cite{fanon2008black}.}~\cite{aktar2024indigeneity}. In such complex dynamics around people's civic belonging in nation-states, Stokke's open-ended analytical framework of citizenship is useful~\cite{stokke2017politics}, where he proposed four interconnected dimensions: legal status, membership, rights, and participation. Membership and legal status refer to cultural and judicial inclusion in communities, whereas rights and participation refer to the entitlements and responsibilities that follow. These four dimensions are mutually constitutive and represent different entry points and potential priorities in the politics and structure of civic inclusion and power.

\subsection{Interaction and Moderation as Platform Governance}

Early social computing research often studied online platforms through the notion of ``public spheres"~\cite{semaan2015designing, semaan2015navigating, papacharissi2002virtual}, where citizens come together, exchange opinions regarding public affairs, discuss, deliberate, and eventually form public opinion~\cite{habermas2020public}. While researchers have also critiqued the associated utopian assumptions considering the politics around interaction~\cite{lampe2014crowdsourcing}, moderation~\cite{gillespie2018custodians}, and algorithms~\cite{geiger2009does, geiger2016bot}, scholars have found such civics-oriented and associated metaphors-based approaches to examine platform governance~\cite{fan2020digital, seering2022metaphors}. But what does such governance do?

Platform governance, exercised through content moderation, community participation, and various sociotechnical mechanisms, aims to create safer and more inclusive online environments~\cite{gillespie2018custodians}, maintain community standards~\cite{seering2022metaphors}, protect users from harmful content~\cite{scheuerman2021framework}, and ensure legal compliance~\cite{fiesler2023chilling}. However, this involves trade-offs, where moderators balance free speech with user safety, negotiate among the interests of users, stakeholders, and regulatory requirements, and establish social and algorithmic mechanisms for various levels of efficiency, accountability, and transparency~\cite{jiang2023trade}. Challenges in such trade-offs often emerge, especially from the varied acceptability of and dominant beliefs around certain topics in different cultures and backgrounds~\cite{jiang2021understanding}.

When certain identities and values are consistently prioritized over others by governance practices, it shapes the platform's identity~\cite{das2021jol, gilbert2020run}. People's identities across race, gender, ethnicity, religion, nationality, linguistic norms, sexual orientation, etc., become the dimensions of such discrimination and marginalization. For example, Gilbert found that ``the default masculine whiteness of Reddit" marginalizes women, Black, and LGBTQ+ people~\cite{gilbert2020run}. In studying the experience of BnQuora users from different religions and nationalities within a single ethnolinguistic group (i.e., the Bengali people) with platform governance, Das and colleagues described how centralized and collective surveillance by content moderators and majority users, respectively, control the users' interaction leading minority users to self-surveil their interaction and self-imprison themselves~\cite{das2021jol}.

Similarly, many recent works in ICTD and social computing have started exploring marginalized users' experience with platform governance~\cite{sultana2022toleration, ma2023users}. While researchers have developed various computational approaches to mitigate the issues with harmful online content~\cite{rifat2024combating, koshy2023measuring} and evaluated the feasibility of using algorithms instead of or besides humans in content moderation~\cite{das2024colonial, molina2022ai} and conceptualized various frameworks for that~\cite{sun2022design, vaidya2021conceptualizing}. However, a deep cultural understanding of tolerance, civility, decorum, trust, and the history of the corresponding communities should be central to approaches to this socio-technical problem~\cite{sultana2022toleration, das2021jol}. As moderation policies often overlook different intersectionalities (e.g., considering countries as proxies for cultures, under-representation of minorities in moderation bodies), the mismatch between moderators' and users' values and perspectives often leads to the erasure of socially marginalized voices online~\cite{das2021jol, thach2024trans}. Besides the lack of moderators' efforts in supporting intersectionally marginalized communities, users from majority groups often run coordinated efforts to influence, disrupt, or manipulate online content or discussions, which researchers dubbed as ``brigading"~\cite{dosono2020decolonizing}. This often results in skewed representation, increased harassment, and sectarian incitement~\cite{dash2022insights}. In the context of the Global South, researchers have studied users' experience with and on various online platforms through the lens of algorithmic coloniality~\cite{das2021jol} or an extension of historical colonization~\cite{shahid2023decolonizing}.

Similar to the analogies drawn between the practices of historical colonialism and platform governance, researchers have also used metaphors to understand moderators' roles and governance structures in online platforms~\cite{seering2022metaphors, fan2020digital}. Based on various roles, such as nurturing and supporting, overseeing and facilitating, and governing and regulating communities, Seering et al.~\cite{seering2022metaphors} found users to describe a moderator as custodian, police, dictator, governor, referee, manager, etc. To facilitate users' democratic participation and trust in platform governance's legitimacy, researchers proposed emulating legal and civics-oriented approaches like jury decision-making, blind voting, modularity, and deliberation for adjudicating content moderation cases~\cite{fan2020digital, schneider2021modular, lenhart2024contentr}.

While these prior works are useful in understanding governance practices, roles, and structures from the content moderators' side, in this work, we are adopting civic-oriented perspectives to understand subaltern users' experience with platform governance in postcolonial contexts.
\section{Methods}\label{sec:methods}
This paper follows a series of studies that focused on the experience of users from marginalized communities on online platforms in postcolonial contexts. While the intersection of race and gender has received much attention in social computing and ICTD research, in this particular paper, we are interested in exploring how, within the same racial and colonially marginalized context, the intersectional ethnic minority and gender identities shape users' experience on online platforms in Bangladesh. By examining their encounters with hate speech in particular, we want to understand their legal standing, sense of cultural belonging, ability to express opinions and access opportunities, and extent of resistance to hate speech through active participation in content moderation.

\subsection{Recruitment and Semi-structured Interviews}
Toward our research goal, we conducted a qualitative study through semi-structured interviews. While little research focused on Bangladeshi ethnic minorities' experience with ICT platforms, prior research has highlighted various unique sociotechnical challenges women face in terms of access and toxic interaction (e.g., hate speech)~\cite{sultana2018design, sultana2021opaque, nova2019online}. We contacted potential participants who identified as ethnic minorities and women through a combination of convenience, purposive, and snowball samplings. We got approval from X University in Bangladesh for our study materials (e.g., questionnaire, protocol) before beginning recruitment. We interviewed 18 participants (17 female and 1 male) between March 2024 and April 2024. Among them, participants came from Chakma (5), Garo (2), Tripura (2), and Bengali (9) ethnic groups. Interviewing participants from both ethnic majority Bengali communities and ethnic minority groups (e.g., Chakma) helped us distinguish the challenges that are intersectional rather than the ones stemming from either gender or ethnicity.

Participation in the study was entirely voluntary. The first author read out the oral consent form to the participants, who all orally gave permission to record the interviews. We asked the participants about their demographic backgrounds, encounters with hate speech on online platforms, the impacts of such encounters on their interactions, and their perceptions of and responses through different local policies, social strategies, and technical processes. We transcribed the interview recordings, anonymized them, and translated them into English prior to data analysis.

\subsection{Data Analysis}
This paper used a dual inductive and deductive approach to data analysis. Such an approach is commonly used in understanding people's practices around technologies in the Global South~\cite{das2022collaborative, kumar2018uber, rifat2024politics}. First, we identified the open codes, i.e., abstract representations of the entities, concepts, and interactions that repeatedly appeared in the interview transcripts through qualitative analysis. Some examples of inductive themes are: ``stigmatizing Indigenous ethnic minorities' practices," ``physical appearance of ethnic minorities," and ``fetishizing Indigenous women." Given the contextual nature of interview data, we did not calculate an inter-rater reliability score~\cite{mcdonald2019reliability}. Based on the patterns that emerged, we noted the relationship between the intersection of ethnicity, gender, and concerns around membership, legal status, rights, and participation. Upon reviewing social computing research on online communities, we identified that metaphors in content moderation, such as the civics-oriented ones, were effective for structured understanding of these concerns online. Hence, we chose Stokke's analytical framework of the politics of citizenship~\cite{stokke2017politics} to examine intersectionally marginalized users' juridical and cultural inclusion, rights, and participation.

\subsection{Positionality Statement}
Prior research has highlighted how the researchers’ identities may reflexively address certain tensions and bring affinities into perspective in studying marginalized communities~\cite{schlesinger2017intersectional, liang2021embracing}. All the authors (two women and three men) of this paper were born and raised in Bangladesh and are members of the Bengali ethnolinguistic group. Four of them are Muslims, and one identifies as a Hindu belonging to an underprivileged non-Aryan aboriginal caste. In addition to their varied sociocultural perspectives, all authors' backgrounds in computer science, with two authors' prior scholarship related to decolonial, postcolonial, marginalized identity, and social computing research, have informed and guided the motivation and execution of this study.

\subsection{Limitations and Future Work}
Due to most of our participants being recruited at or through social connections through a university, the major limitation of our study is the limited socioeconomic and age diversity of our participants (18-28 years). While this restricts the generalizability of the findings, within the qualitative tradition of research~\cite{leung2015validity}, the goal of this work is not to produce generalizability but rather to study a specific phenomenon in a focused context. In the future, we plan to include a more diverse set of participants. A follow-up of this study will aim to develop socio-culturally aware algorithmic and policy-driven approaches to mitigate online hate and toxicity targeted toward intersectionally marginalized communities.
\section{Results}\label{sec:results}
In this section, we will discuss our empirical findings by drawing analogies with the components of citizenship as defined by Stokke~\cite{stokke2017politics}. Here, we examine the users' legal status as their juridical inclusion, membership as their cultural inclusion, rights as their entitlement to voice, and participation as their platform responsibilities.

\subsection{Legal Status: Users' Juridical Inclusion}
Who is considered a Bangladeshi user when using online platforms? An individual is considered a user on a platform simply by joining it. Civics-oriented analogies can be drawn between this process of becoming a user and the process of becoming a citizen in modern states. While citizenship is acquired based on the citizenship of parents (\textit{jus sanguinis}), being born within the territory of a state (\textit{jus soli}), marrying a citizen (\textit{jus matrimonii}), or residence for a given period (\textit{jus domicile})~\cite{stokke2017politics}, computing researchers have also discussed citizenship informationalized through technology~\cite{cheney2017we}. For example, US federal bodies like the National Security Agency (NSA) formally rely on \textit{jus algorithmi}--a formal ``citizenship" reconceptualized based on people's data~\cite{cheney2017we}. By virtue of using contemporary online platforms, i.e., \textit{jus usus}, most of which have some algorithmic components to those, individuals become subject to the arbitration of \textit{jus algorithmi}. P1 describes its pervasiveness,

\begin{quote}
    Social media is free for everyone, so many people who are illiterate on social media can use the platform, so their thinking is different. So, to avoid bad comments or bullying, it is better to add some restrictions. \hfill(P1, male, Chakma)
\end{quote}

This statement indicates the hierarchies among users. This entails stratified possibilities, jurisdictions, and outcomes for their presence on online platforms. So, who should design and implement these restrictions, and for whom? Beyond platform policies--which are a contractual relation between users and the platforms, both users and online spaces are also subject to local laws. But how are the users' juridical inclusion determined? Unlike its \textit{soli} or \textit{sanguinis} counterparts, informationalized citizenships (e.g., \textit{jus usus}, \textit{jus algoritmi}) are not something to be proved once and for all but continuously performed for algorithmic arbiters. This blurs the idealized image of a binary distinction between citizens and non-citizens and of equality among all users. For example, to determine how local laws apply to a user, without specific information, the NSA presumes a user to be a foreigner until positively identified as a US person. As the understanding of hate speech significantly varies across different cultures~\cite{jiang2021understanding}, national and state governments often develop legal articles to make provisions for the trial of offenses committed through digital platforms, such as the Digital Security Act (DSA) in Bangladesh. Contrary to NSA's approach that differentiates between US and non-US persons~\cite{cheney2017we}, Bangladesh's DSA claims provision over offenses by any person or institution in and outside of the country. DSA and its predecessor Section 57 of the ICT Act~\cite{tbs2023deaths} have been criticized for disproportionate prosecution of minority users~\cite{das2023blasphemy}. While users from the Global South are legally vulnerable in general~\cite{ahmed2017privacy, arora2019general, haque2020privacy}, their subaltern identities (e.g., minority) often lead to intersectional marginalization and experience with hate speech.

\subsection{Membership: Users' Cultural Inclusion}
While legal identification of users (e.g., how and which local laws should apply to them) often relies on the territories from where they access online platforms, it is fundamentally challenged by cultural diversity and identity politics~\cite{stokke2017politics}, where the notion of users' membership in local communities draws a distinction between insiders and outsiders. We need to examine whether and to what extent subaltern users are considered to be members of Bangladeshi online communities. Drawing an analogy to the civics framework of citizenship, the concern around membership inherently becomes a question of cultural inclusion. In the case of nationhood, scholars highlight a basic distinction between ethnocultural and juridical-political constructions of nationhood~\cite{stokke2017politics}. For example, whereas some national communities are built around a cultural essence (e.g., German nationhood being based on a notion of an ethnic community with strong ties to a historical homeland), other nations are defined through a territorial state formation (e.g., French nationhood revolving around people living under common law and the same legislative assembly within the territorial state)~\cite{brubaker2009citizenship}. In Bangladesh, while 98\% of its people are Bengali, there are many ethnic minority communities in the country. As computing systems tend to rely heavily on reductionist representation~\cite{dourish2012ubicomp}, the users from Bangladesh who identify as ethnic minorities and women would have similar \textit{jus algorithmi} or \textit{jus usus} for an online platform based on their locations. However, similar to Bengali female users' varied experiences from that of Bengali male users, highlighted by prior work~\cite{nova2019online, nova2018silenced}, our study highlights how the ethnic minorities' cultural diversities with the majority Bengali communities shape their online experience with hate speech.

Our participants from Bangladeshi ethnic minority communities described how they are identified and made targets of hate speech based on their physical appearance. P1 described one such incident:

\begin{quote}
    Once, when I was a college student, some of my friends and I posted photos on Facebook. One of my Facebook friends made a comment which hurt me a lot. You know I am from a minority community, and others are Bengali. In the photo, I was at a side. After posting it, my Facebook friend commented, ``Why did you keep her at a side? ... Her appearance is different. ... She is from a minority community." It was very insulting. \hfill (P1, male, Chakma)
\end{quote}

Here, users from majority communities, even the ones our participants considered part of their social circles, were alienating them based on their different ethnicity. Our female participants from the Bangladeshi Bengali communities also described experiences of body shaming. Participant P11 shared an acquaintance's story:

\begin{quote}
    The incident happened not with me but with a known senior sister\footnote{In Bangladeshi culture, senior female friends are often referred to as apa/didi, which means senior sister, though they are usually not related by blood.} from Chittagong. She is a businesswoman. Once, she was showing dresses for her business. She was a bit healthy (colloquially modest way to describe someone who is overweight), so when she was in real life, many people commented in bad taste, like ``Why are you so healthy? Why do you look odd?"  [...] Body shaming, basically. She was married, and it was so insulting for her as some men started bullying her. \hfill (P11, female, Bengali)
\end{quote}

In contrast to the incident described above, where users from the majority community attempted to distance the minority users, our participants also described cases where they were tokenized and exoticized. Participant P5 described her discomfort with such interaction:

\begin{quote}
    With me, it was not bullying, but it was like a random man actually sending a friend request mentioning that he was interested in Chakma\footnote{One of the largest ethnic minority groups in Bangladesh}. So, do you want to be my Facebook friend? This is a common problem, and my friends and sisters always face this issue. \hfill(P5, female, Tripura)
\end{quote}

This statement highlights how female users from ethnic minority communities in Bangladesh are reduced to their ethnicity for sensual desires and superficially included for appearance's sake. Similar to how subaltern users' interaction is influenced by their physical appearance being different from that of the majority user communities (e.g., Bengali men), their cultural practices are often stigmatized in relation to normative Bangladeshi Bengali practices and viewed through a lens of ill-informed stereotypes. Beyond physical appearance, our participants described how Adivasi communities often become subject to racial slurs based on their languages and accents. P5 shared:

\begin{quote}
    I know a Chakma senior sister who has a Facebook page [...], where someone made some bad comments like ``ching chong." Moreover, celebrity Chakma girls always get racist comments, which is embarrassing for them. \hfill(P5, female, Tripura)
\end{quote}

These hateful comments our participants and their acquaintances received demonstrate the othering of Adivasi and ethnic minorities in Bangladeshi online communities. The concerns of subaltern users truly being members of these communities in terms of cultural inclusion demand an understanding of normative and non-normative cultural practices in the country. For example, while the nationhood of Bangladesh is intertwined with Bengali ethnolinguistic identity and the country's constitution upholds secularism as one of the fundamental principles~\cite{bangladesh1972constitution}, more than 91\% of the population being Muslim~\cite{bsb2022preliminary} and the state religion being Islam, Islamic values strongly influence and shape the cultural norms (e.g., food habit) of the country.

Let's consider participant P5's conversation with someone from Bangladeshi Bengali community:

\begin{quote}
    A friend asked me, ``Is alcohol your main food?" I was shocked and asked where she knew that. She told me that from the class four textbook. ... I think this is not only insulting but also humiliating the tribal people of Bangladesh. \hfill(P5, female, Tripura)
\end{quote}

\begin{figure}[!ht]
    \centering
    \includegraphics[width=0.5\textwidth]{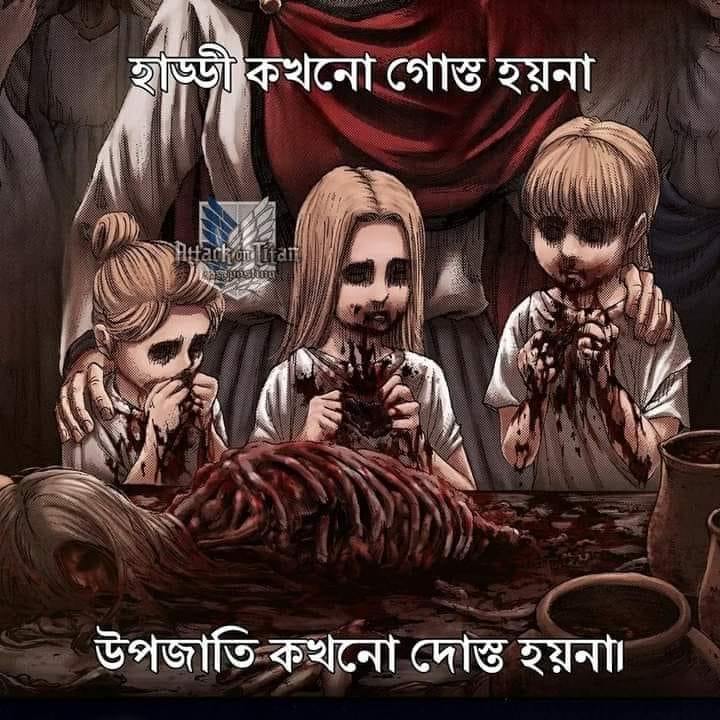}
    \caption{A viral photo card was spreading hatred against Adivasi communities. Rhyming Bengali text in it translates as ``Bones never become flesh and \textit{upajati} (meaning and use explained in section~\ref{sec:literature_review}) never become friends." We do not provide the URL to abstain from contributing to popularizing the source of such hateful content.}
    \label{fig:adivasi_cannibal_rumor}
\end{figure}

Views like alcoholic drinks being the main food of Adivasi communities, a semantically incorrect statement to begin with, are largely driven by stereotypes about non-normative cultures and food habits. Though most Adivasi communities in the Chittagong Hill Tracts (CHT) region of Bangladesh belong to non-Muslim faiths~\cite{minority2018adivasis} and can consume alcohol with a legal permit~\cite{woods2022bangladesh}, its consumption in Bangladesh is severely stigmatized and considered non-normative in Bangladesh owing to the strong influence of the religious values of the majority communities. As national curriculum and textbooks often shape people's general knowledge of diverse communities~\cite{das2024reimagining}, our participants emphasized that the authorities should consult corresponding minority community experts while including and narrating such information. Several of our participants referred to a recent incident that gained a lot of attention in mainstream and social media: a few Adivasi men kidnapped and murdered a Bengali college student. Afterward, fake stories, photo cards (see example in Figure~\ref{fig:adivasi_cannibal_rumor}), and videos on Bangladeshi social media started claiming the offender had ``cooked and eaten" the victim's flesh and spread rumors of the Adivasi communities living in the CHT to be cannibals.

Participant P4 described how the lack of cultural sensitivities and malignant stereotypes weaponized the incident to demonize the ethnic minorities at large:

\begin{quote}
    Recently, there has been a discussion about minority groups. For example, aboriginal populations are cannibalistic. So, for joking, friends say things like \textit{``Tora to manus khash"} (meaning, ``You guys eat human flesh"). Even if it is a joke, it hurts me. \hfill(P4, female, Tripura)
\end{quote}

Characterizing ``subaltern" as a negative space from where one cannot express their concerns and opinions, Spivak rhetorically asked~\cite{spivak2003can}, ``Can the subaltern speak?" Similarly, the question here is whether these subaltern users (e.g., Adivasis and women)--intersectionally marginalized users in the Global South can make their voices heard while facing hateful speech on online platforms.

\subsection{Rights: Users' Entitlement to ``Voice"}
People are entitled to a set of rights as citizens of modern nation-states, usually understood through a threefold typology of civil, political, and social rights. Among these, civil rights protect freedom of conscience and choice (e.g., free speech and press, and freedom of religion), political rights capacitate people to express opposition and to protest, and social rights include enabling opportunity (e.g., in education and the labor market)~\cite{stokke2017politics}. While these rights have various complexities in terms of citizenship in a state, our work builds an analogy in the case of users in online platforms. We are inspired by the notion of sociomateriality in postcolonial contexts~\cite{das2022collaborative, das2024reimagining, das2022understanding} that highlights the importance and effectiveness of online discourse in real-life movements for sociopolitical rights. Prior ICTD scholarship conceptualizes the ability to express people's rightful opinion through technology as ``voice"~\cite{ahmed2018designing}. Subject to the politics of access, autonomy, and accountability~\cite{ahmed2018designing}, technologies like online platforms augment, shape, and impede users' entitlement to voice their sociopolitical concerns~\cite{das2022understanding} and demands for various rights~\cite{das2022collaborative}. Therefore, a civics-oriented approach to understanding subaltern users' experience in Bangladeshi social media needs to examine to what extent users can voice themselves online and ask for rights in those spaces.

While prior social computing researchers described content moderation (e.g., determining whether a post is harmful) as subject to trade-off among various factors and values~\cite{jiang2023trade, jiang2021understanding}, our study drawing an analogy between the users' rights in online platforms and rights associated with citizenship highlights tensions between different kinds of rights. For example, reflective discussions that negotiate between principles of universality and equality and efforts that aim at addressing inequalities between social groups often surface in Bangladeshi social media, especially in the context of various provisions of the CHT peace accord and reservation for underprivileged ethnic minority communities. While some of these discourses are contextual and demand careful ethical and legal consideration, some experiences of our participants were clearly shaped by majoritarianism. As participant P3 shared:

\begin{quote}
    I never post anything odd that can trigger hate speech. ... I have seen it in the case of my friends. Recently, one of my university Bangali friends posted on Facebook, wondering about considering public holidays for ethnic minority groups' different occasions like Bizhu, Baisabi, etc. In that post, I found a mutual friend's comment that minorities should not be eligible for those holidays. ... Such a narrow-minded comment from a person in an independent country looks odd. Later, she deleted the post. \hfill (P3, female, Garo)
\end{quote}

Like Participant P3, many of our participants talked about self-suppressing their voices in fear of further hate speech. Despite being interested in creating Facebook pages to be active, which often helps in gaining and sustaining popularity in Bangladeshi social media~\cite{das2022understanding}, they often lock their profile and do not engage in ``social browsing"~\cite{lampe2006face} (e.g., accepting unknown people's friend requests). This shows how majoritarian domination violates subaltern users' freedom of speech. In the case described above, a user from the majority community came forward to raise their voice in support of the ethnic minorities. She was talking about the new year festivals like Bizhu and Baisabi that are observed through various cultural and religious rituals by Adivasi communities\footnote{It is known as Bizhu among the Chakmas, Sangrai among the Marmas, and Baisab among the Tripuras.}. Whereas digital activism is often championed in ICTD literature, our study shows how subaltern users, like those from adivasi communities, cannot exercise their civil rights like freedom of speech or demanding freedom of religion on online platforms and how the support of the allies from majority communities also fall short in face of severe majoritarianism.

Similarly, our participants discussed how their efforts to voice their political rights are barred in online communities. For example, as previously discussed in section~\ref{sec:literature_review}, adivasi women's sexual abuse has been a major concern with the historically state-sponsored Bengali settlements in the CHT region. With the offenders of many of these incidents going unpunished, the Adivasi users sometimes demand their rights to justice online. Participant P2 said:

\begin{quote}
    I posted on Facebook about a rape incident [of an indigenous woman] in Bandarban in 2012 and accused the army. I still didn't know what actually happened. Then, many people made bad comments. They accused minority groups that [the victim] may have done that for money. It was very insulting to all minority people in the area. \hfill (P2, female, Garo)
\end{quote}

While the participant speculated about the army's involvement in the incident, many users from the majority Bengali communities commenting on her post blamed the victim and her community unfairly. Related to the juridical exclusion and cultural alienation of Adivasi communities that we discussed in the previous two subsections, this broadly highlights the futility of online platforms in voicing subaltern users' political rights, whereas prior works have found the same platforms to be effective in Bengali people's sociopolitical discourse~\cite{das2022understanding}. Similarly, whereas social media has opened new avenues for engaging in the labor market through home-based businesses in Bangladesh~\cite{mim2022f}, subaltern users are often deterred from accessing these opportunities. For example, participant P11 described how a female entrepreneur missed out on business opportunities for online hate speech:

\begin{quote}
    Due to body shaming, first, she quit her business life, but after three months, she started again and then replied to them, asking why they were body shaming her and What her fault was if she looked like that. \hfill(P11, female, Bengali)
\end{quote}

While appealing to peers in this way often becomes an effective way to voice one's concerns, a well-functioning governance mechanism, whether in state or online platforms, should give ways for its subjects -- citizens or users -- to formally bring those concerns to its attention.

\subsection{Participation: Users' Platform Responsibilities}
While a good citizen is viewed as a self-governing member of the state, a civics-oriented communitarian perspective of users emphasizes their participation at the community level in an online platform. That means users also have responsibilities to be involved in the governance and content moderation of online platforms. While most platforms are inclined toward either a centralized or distributed approach, they often strike a balance between these extremities. In doing so, some users engage in extensive participation, such as the moderators and administrators who can devote time to the duties of platform governance. In addition to surveillance of such a centralized moderation body, they also rely on users' collective involvement in decision-making and monitoring of content. However, as previously found by prior works, such collective participation of users in cleansing the platform of hate speech relies on the strength of numbers. Participant P5 explained:

\begin{quote}
    Facebook doesn't delete any ID for one report--it requires many [reports against one ID]. Sometimes, despite a report, it doesn't delete the ID. I like to report different types of comments. Recently, I made a report, but Facebook didn't take any action. This is why minority community women who are bullied are demotivated to report those IDs, and such incidents are increasing day by day. \hfill(P5, female, Tripura)
\end{quote}

While online platform moderation systems' reliance on the number of reports makes sense in certain cases, the similar quantitative threshold of collective reports is unfair and disproportionate for subaltern users like women from Indigenous communities. The problem with reliance on collective reporting in platform governance can also stem from varied levels of user participation. In civics literature, citizens are divided into three types: participant, non-participant, and opportunistic~\cite{stokke2017politics}. Building on that, we can identify users in online platforms who are either incorporated, sometimes commercially, as community managers by the platform itself or the ones who actively participate in reporting harmful content they encounter. Non-participating users are the ones who accept authorities of platform governance and justify their cynicism and inactivity with the impossibility of achieving real change in the platform's moderation practices or the subaltern users, like the ones participant P5 mentioned in the statement above, who are alienated through the lack of required resources and, in the example statement above, support from other users. There also exist opportunistic users who prioritize their own interests and participate in content moderation work, such as reporting only if their participation would directly affect their interests. Participant and opportunist users adopt various strategies in platform governance. Participant P8 explained:

\begin{quote}
    When a woman posts something, 5-6 men become a group and then start making bad comments, one after another. ... She first tries to reply to those bullies. The victim then shares those [bullies'] Facebook IDs with her friends to report. But when the woman replies to those bullies, and his friends support that boy, then it becomes tough to report all that is. Since reporting IDs for a bully is a lengthy process, she ignores it. \hfill(P8, female, Bengali)
\end{quote}

This example captures the process of brigading. The collective targeted attack by the bullies, who are opportunistic users in this case, on a female user online demotivates the female user to report bullying to the platform governance and turns her into a non-participant user. Participant P5 mentioned that even when the victim's friends support them in the cases of brigading, it often leads to a conflict between two groups without any possibility of better interaction in the future.

Though online communities are often described as participatory spaces, like in the cases of citizens in a state~\cite{holston2007insurgent}, targeted groups are invited to participate on terms that are defined top-down or through norms of the majority~\cite{das2021jol}. Here, rather than being sovereign individuals who express themselves through democratic participation, the subjects are invited to participate on terms set by the state or, in this paper's case, online platforms. Let's consider what Participant P12 said about Facebook's reporting system:

\begin{quote}
    A few options are available. I myself have done some reports. ... Yes, bullying is available, but options are very few or limited. ... There should be more described options, and for what reason we are reporting, there should be adequate spaces for it to be mentioned. \hfill(P12, female, Bengali)
\end{quote}

Here, the participant described how users participate in content moderation on Facebook according to the terms or high-level options set by the platform, which often do not capture the nuanced contexts of hate speech or harmful content the subaltern users encounter. Similar to the participant above, who asked for spaces to explain the particular situations with each reporting, Participant P1 talked about a need for ways in the reporting system for the users to directly add options to the current list of categories.
\section{Discussion}\label{sec:discussion}

Understanding the subaltern users' experience with hate speech through a civics-oriented approach helps us develop the idea of ``usership". Analogous to the concept of citizenship, usership highlights the spectrum of legal status, membership, rights, and participation of users in online platforms. In this section, we discuss the concept of ``usership" based on this analogy, reflect on the limitations inherent to such a metaphor-based framework, and outline the implications for ICTD practitioners and platform designers.

\subsection{Usership as a Conceptual Framework}
\begin{figure}[!ht]
    \centering
    \includegraphics[width=\textwidth]{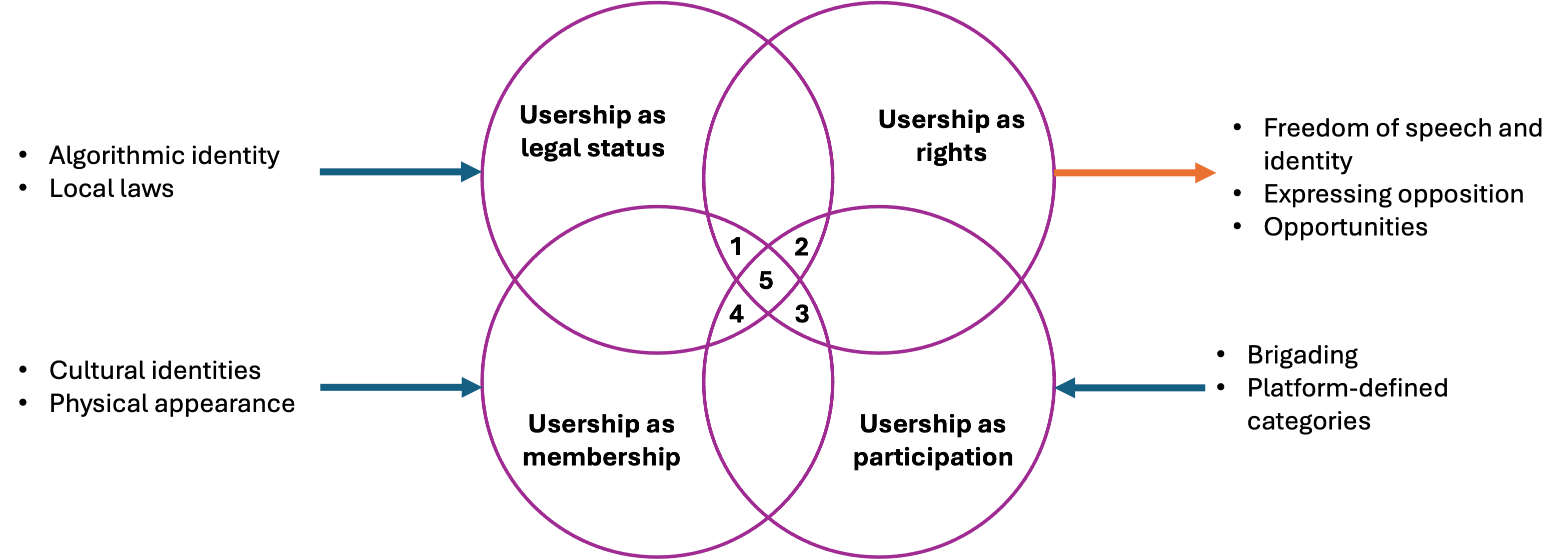}
    \caption{Dimensions and stratification of usership}
    \label{fig:usership_dimension}
\end{figure}

By adopting and modifying Stokke's framework of citizenship~\cite{stokke2017politics}, we visualized the dimensions and stratification of usership in Figure~\ref{fig:usership_dimension}. When users join and use a platform, their contract with the platform--what they can and are expected to do is characterized by \textit{jus usus} (status by the right of using). In addition, the users' legal status is also determined by the local laws in their geographic location and that of the platform, which often relies on the users' algorithmic identity, which we explained through the concept of \textit{jus algorithmii} (status by the right of algorithms). Here, the political factors (e.g., local laws) and identity constructed by sociotechnical systems (e.g., algorithms) influence the juridical inclusion of users, i.e., the legal status of usership.

Our results showed how ethnic minority users are excluded from Bangladeshi online communities. Their alienation is often based on different physical appearances, stereotypes, ill-informed rumors, languages, and cultural practices. The study highlights how intersectional marginalization, such as in the form of exclusion and exoticization, of Adivasi women is unique to them and cannot be captured by just studying Bangladeshi Adivasi or Bangladeshi female users. This foregrounds how societal factors like norms around looks, food, and practices shape users' cultural inclusion, in other words, their membership in online communities. As social group identities influence users' membership in online platforms, our study demonstrated how that impacts subaltern users' interaction and their ability to benefit from content moderation processes (e.g., reporting). The study showed how brigading of opportunist users, for example, those from majority communities and dominant gender identity, as well as top-down defined terms and hate speech categories from the platform, limit subaltern users' participation in the collective platform governance, such as through identifying and cleansing harmful content.

To understand how online platforms in Bangladesh shape subaltern users' rights, we drew on the concept of voice. This idea is central to subaltern postcolonial studies~\cite{spivak2003can} and one of the major threads in ICTD scholarship~\cite{ahmed2018designing}, which focuses on concerns around access, autonomy, and accountability. By closely examining Bangladeshi subaltern users' freedom of speech, identity (e.g., religious) expression, ability to protest and convey opposition in political matters, and access to opportunities (e.g., commercial) on online platforms through this concept, we demonstrated how these platforms differentially contribute to users' entitlements to civil, social, and political rights.

While ideally, one's usership should be at the intersection of all four dimensions, of course, the dimensions of usership overlap in various ways to stratify the users of a platform. For example, stratum 1 in Figure~\ref{fig:usership_dimension} indicates the ones who are considered users by the platform under particular local legal authorities, culturally belong to the community, and enjoy certain rights but cannot participate in content moderation. In most cases, this mostly includes the general users who are not moderators of any online spaces or do not have the privilege to report harmful content or participate in meta-discussion boards (e.g., users with low reputation points in StackExchange platforms). Similarly, stratum 2 contains users who meet the legal requirements of the platform and local laws, can exercise rights, and actively participate but are not culturally integrated with the online community. Examples of these users can be new users in a subreddit unfamiliar with the community norms. While platforms almost always have jurisdiction over their users, a user (in stratum 3) might not have legal status in terms of local laws despite having cultural membership, certain rights, and participation (e.g., minor users). Similarly, stratum 4 covers the users who have legal status, membership, and ability to participate but without rights. For example, a YouTuber who cannot monetize their videos but can appeal the platform's decision in that regard. There can also be situations where two or only one dimension of usership is satisfied.

\subsection{Shortcomings of the Usership Framework}
While the analogy between citizenship and usership provides a structured lens through which we can examine user interactions on online platforms, its limitations also must be considered in an ICTD context. Central differences in citizenship and usership emerge from their frequency/possibility of changing and the nature of governance between the entities involved: nation-states and corporate platforms. Citizenship is often tied to relatively fixed identities—such as nationality or legal status—embedded in institutional frameworks and difficult to change. In contrast, usership on online platforms is marked by more fluid and dynamic identities, as users can adopt multiple personas, remain anonymous, or shift between roles across different platforms. This flexibility is not easily captured by the relatively rigid concept of citizenship, where identity is more static.

Furthermore, Citizenship typically involves a legal and political relationship between individuals and a sovereign state, which provides rights, responsibilities, and protections within a framework of laws and political structures designed to balance individual and collective interests. However, online platforms are corporate entities driven primarily by business objectives, such as profit maximization, user engagement, and operational efficiency. Governance in these contexts often prioritizes these business interests over the protection of user rights, which introduces complexities when applying the citizenship analogy to usership. At the same time, when the values of these corporate bodies do not align with those of the sovereign state where they operate, it poses ethical dilemmas and legal complexities around users. As such, the analogy between usership and citizenship may obscure important differences in how individuals navigate identity, agency, and participation in digital environments. These distinctions suggest that while the citizenship framework offers valuable insights, it may not fully account for the fluidity and complexity of usership in online spaces.

\subsection{Implications of the framework for HCI and ICTD Research}
Altogether, how the dimensions of usership shape and are shaped by users' identity, agency, and sociocultural and legal influences on technology emphasizes how subaltern users interact with and through online platforms and experience the broader socio-technical environment. Criticism of such a conceptualization of usership can emerge from post-modernist approaches in HCI. For example, using ``user" as a proxy construct for ``human," we lose nuances about subject positions. Similarly, adopting a lens of ``citizen" over ``public" can bring certain ontological limitations. However, Brubaker and Baumer argue that a structural notion of user prioritizes efficiency, calculability, and predictability~\cite{baumer2017post}. Especially as many recent works are adopting structural ways to understand content moderation in online platforms, similarly structured approaches to understanding users' experiences would help us get a more comprehensive understanding of those spaces.

Hence, despite these limitations, the analogy between citizenship and usership remains a useful conceptual tool for understanding how users are included, excluded, or stratified online, especially in regions where marginalized communities face systemic exclusion from physical and digital infrastructures. By offering a structured way to analyze the multiple dimensions of usership, such as legal standing, rights, participation, and cultural belonging, the framework serves as an important starting point for exploring how digital environments mediate users' access and experiences.

\section{Conclusion}
This paper examines intersectionally marginalized users' experience with hate speech online. We found how the users in Bangladesh who come from Indigenous ethnic minority communities and/or identify as women face cultural exclusion and cannot access technologically mediated opportunities, exercise sociopolitical rights, and participate in decision-making for adjudicating online content to the same extent as the ethnic majorities and male users. Based on the empirical findings, we developed a civics-oriented framework of ``usership" that helps us understand users' legal status, membership, rights, and participation in platform governance. This framework, informed by the perspectives in the Global South, would provide a user-centered metaphor for studying moderation.

\bibliographystyle{ACM-Reference-Format}
\bibliography{sample-ref}


\begin{thebibliography}{101}


\ifx \showCODEN    \undefined \def \showCODEN     #1{\unskip}     \fi
\ifx \showDOI      \undefined \def \showDOI       #1{#1}\fi
\ifx \showISBNx    \undefined \def \showISBNx     #1{\unskip}     \fi
\ifx \showISBNxiii \undefined \def \showISBNxiii  #1{\unskip}     \fi
\ifx \showISSN     \undefined \def \showISSN      #1{\unskip}     \fi
\ifx \showLCCN     \undefined \def \showLCCN      #1{\unskip}     \fi
\ifx \shownote     \undefined \def \shownote      #1{#1}          \fi
\ifx \showarticletitle \undefined \def \showarticletitle #1{#1}   \fi
\ifx \showURL      \undefined \def \showURL       {\relax}        \fi
\providecommand\bibfield[2]{#2}
\providecommand\bibinfo[2]{#2}
\providecommand\natexlab[1]{#1}
\providecommand\showeprint[2][]{arXiv:#2}

\bibitem[Ahmed(2018)]%
        {ahmed2018designing}
\bibfield{author}{\bibinfo{person}{Syed~Ishtiaque Ahmed}.}
  \bibinfo{year}{2018}\natexlab{}.
\newblock \emph{\bibinfo{title}{Designing for voice: The challenges of access,
  autonomy, and accountability}}.
\newblock \bibinfo{thesistype}{Ph.\,D. Dissertation}. \bibinfo{school}{Cornell
  University}.
\newblock


\bibitem[Ahmed et~al\mbox{.}(2017)]%
        {ahmed2017privacy}
\bibfield{author}{\bibinfo{person}{Syed~Ishtiaque Ahmed},
  \bibinfo{person}{Md~Romael Haque}, \bibinfo{person}{Shion Guha},
  \bibinfo{person}{Md~Rashidujjaman Rifat}, {and} \bibinfo{person}{Nicola
  Dell}.} \bibinfo{year}{2017}\natexlab{}.
\newblock \showarticletitle{Privacy, security, and surveillance in the Global
  South: A study of biometric mobile SIM registration in Bangladesh}. In
  \bibinfo{booktitle}{\emph{Proceedings of the 2017 CHI Conference on Human
  Factors in Computing Systems}}. \bibinfo{pages}{906--918}.
\newblock


\bibitem[Aktar(2024)]%
        {aktar2024indigeneity}
\bibfield{author}{\bibinfo{person}{Solnara Aktar}.}
  \bibinfo{year}{2024}\natexlab{}.
\newblock \bibinfo{title}{Indigeneity and Recognition: Ethnic Minority Rights
  in Bangladesh}.
\newblock
\newblock


\bibitem[Ali(2016)]%
        {ali2016brief}
\bibfield{author}{\bibinfo{person}{Syed~Mustafa Ali}.}
  \bibinfo{year}{2016}\natexlab{}.
\newblock \showarticletitle{A brief introduction to decolonial computing}.
\newblock \bibinfo{journal}{\emph{XRDS: Crossroads, The ACM Magazine for
  Students}} \bibinfo{volume}{22}, \bibinfo{number}{4} (\bibinfo{year}{2016}),
  \bibinfo{pages}{16--21}.
\newblock


\bibitem[Arora(2019)]%
        {arora2019general}
\bibfield{author}{\bibinfo{person}{Payal Arora}.}
  \bibinfo{year}{2019}\natexlab{}.
\newblock \showarticletitle{General data protection regulation—A global
  standard? Privacy futures, digital activism, and surveillance cultures in the
  Global South}.
\newblock \bibinfo{journal}{\emph{Surveillance \& Society}}
  \bibinfo{volume}{17}, \bibinfo{number}{5} (\bibinfo{year}{2019}),
  \bibinfo{pages}{717--725}.
\newblock


\bibitem[Banerjee(1998)]%
        {banerjee1998politics}
\bibfield{author}{\bibinfo{person}{Prathama Banerjee}.}
  \bibinfo{year}{1998}\natexlab{}.
\newblock \bibinfo{booktitle}{\emph{The politics of time:'Primitives' and the
  writing of history in Colonial Bengal}}.
\newblock \bibinfo{publisher}{University of London, School of Oriental and
  African Studies (United Kingdom)}.
\newblock


\bibitem[Baumer and Brubaker(2017)]%
        {baumer2017post}
\bibfield{author}{\bibinfo{person}{Eric~PS Baumer} {and} \bibinfo{person}{Jed~R
  Brubaker}.} \bibinfo{year}{2017}\natexlab{}.
\newblock \showarticletitle{Post-userism}. In
  \bibinfo{booktitle}{\emph{Proceedings of the 2017 CHI Conference on Human
  Factors in Computing Systems}}. \bibinfo{pages}{6291--6303}.
\newblock


\bibitem[Baviskar(2013)]%
        {baviskar2013politics}
\bibfield{author}{\bibinfo{person}{Amita Baviskar}.}
  \bibinfo{year}{2013}\natexlab{}.
\newblock \showarticletitle{The Politics of Being ``Indigenous”}.
\newblock In \bibinfo{booktitle}{\emph{Indigeneity in India}}.
  \bibinfo{publisher}{Routledge}, \bibinfo{pages}{41--58}.
\newblock


\bibitem[Bhatia and Elswah(2024)]%
        {bhatia2024context}
\bibfield{author}{\bibinfo{person}{Aliya Bhatia} {and} \bibinfo{person}{Mona
  Elswah}.} \bibinfo{year}{2024}\natexlab{}.
\newblock \bibinfo{title}{Context Before Code: Meta’s Oversight Board Policy
  Advisory Opinion on the Word “Shaheed” Calls for Language and Cultural
  Nuance in Content Moderation}.
\newblock
  \bibinfo{howpublished}{\url{https://cdt.org/insights/context-before-code-metas-oversight-board-policy-advisory-opinion-on-the-word-shaheed-calls-for-language-and-cultural-nuance-in-content-moderation/}}.
\newblock
\newblock
\shownote{Last accessed: Jul 19, 2024}.


\bibitem[Brubaker(2009)]%
        {brubaker2009citizenship}
\bibfield{author}{\bibinfo{person}{Rogers Brubaker}.}
  \bibinfo{year}{2009}\natexlab{}.
\newblock \bibinfo{booktitle}{\emph{Citizenship and nationhood in France and
  Germany}}.
\newblock \bibinfo{publisher}{Harvard University Press}.
\newblock


\bibitem[Bruckman(2006)]%
        {bruckman2006new}
\bibfield{author}{\bibinfo{person}{Amy Bruckman}.}
  \bibinfo{year}{2006}\natexlab{}.
\newblock \showarticletitle{A new perspective on" community" and its
  implications for computer-mediated communication systems}. In
  \bibinfo{booktitle}{\emph{CHI'06 extended abstracts on Human factors in
  computing systems}}. \bibinfo{pages}{616--621}.
\newblock


\bibitem[BSB(2022)]%
        {bsb2022preliminary}
\bibfield{author}{\bibinfo{person}{Bangladesh Statistics~Bureau BSB}.}
  \bibinfo{year}{2022}\natexlab{}.
\newblock \bibinfo{title}{Preliminary Report on Population and Housing Census
  2022 : English Version}.
\newblock
  \bibinfo{howpublished}{\url{https://sid.portal.gov.bd/sites/default/files/files/sid.portal.gov.bd/publications/01ad1ffe_cfef_4811_af97_594b6c64d7c3/PHC_Preliminary_Report_(English)_August_2022.pdf}}.
\newblock
\newblock
\shownote{[Accessed: Feb 28, 2023]}.


\bibitem[Chakma(2008)]%
        {chakma2008assessing}
\bibfield{author}{\bibinfo{person}{Bhumitra Chakma}.}
  \bibinfo{year}{2008}\natexlab{}.
\newblock \showarticletitle{Assessing the 1997 Chittagong hill tracts peace
  accord}.
\newblock \bibinfo{journal}{\emph{Asian Profile}} \bibinfo{volume}{36},
  \bibinfo{number}{1} (\bibinfo{year}{2008}), \bibinfo{pages}{93}.
\newblock


\bibitem[Chakma(2010)]%
        {chakma2010post}
\bibfield{author}{\bibinfo{person}{Bhumitra Chakma}.}
  \bibinfo{year}{2010}\natexlab{}.
\newblock \showarticletitle{The post-colonial state and minorities: ethnocide
  in the Chittagong Hill Tracts, Bangladesh}.
\newblock \bibinfo{journal}{\emph{Commonwealth \& comparative politics}}
  \bibinfo{volume}{48}, \bibinfo{number}{3} (\bibinfo{year}{2010}),
  \bibinfo{pages}{281--300}.
\newblock


\bibitem[Cheney-Lippold(2017)]%
        {cheney2017we}
\bibfield{author}{\bibinfo{person}{John Cheney-Lippold}.}
  \bibinfo{year}{2017}\natexlab{}.
\newblock \bibinfo{booktitle}{\emph{We are data: Algorithms and the making of
  our digital selves}}.
\newblock \bibinfo{publisher}{New York University Press}.
\newblock


\bibitem[Crenshaw(2013a)]%
        {crenshaw2013demarginalizing}
\bibfield{author}{\bibinfo{person}{Kimberl{\'e} Crenshaw}.}
  \bibinfo{year}{2013}\natexlab{a}.
\newblock \showarticletitle{Demarginalizing the intersection of race and sex: A
  black feminist critique of antidiscrimination doctrine, feminist theory and
  antiracist politics}.
\newblock In \bibinfo{booktitle}{\emph{Feminist legal theories}}.
  \bibinfo{publisher}{Routledge}, \bibinfo{pages}{23--51}.
\newblock


\bibitem[Crenshaw(2013b)]%
        {crenshaw2013mapping}
\bibfield{author}{\bibinfo{person}{Kimberl{\'e}~Williams Crenshaw}.}
  \bibinfo{year}{2013}\natexlab{b}.
\newblock \showarticletitle{Mapping the margins: Intersectionality, identity
  politics, and violence against women of color}.
\newblock In \bibinfo{booktitle}{\emph{The public nature of private violence}}.
  \bibinfo{publisher}{Routledge}, \bibinfo{pages}{93--118}.
\newblock


\bibitem[Das(2023)]%
        {das2023studying}
\bibfield{author}{\bibinfo{person}{Dipto Das}.}
  \bibinfo{year}{2023}\natexlab{}.
\newblock \showarticletitle{Studying Multi-dimensional Marginalization of
  Identity from Decolonial and Postcolonial Perspectives}. In
  \bibinfo{booktitle}{\emph{Companion Publication of the 2023 Conference on
  Computer Supported Cooperative Work and Social Computing}}.
  \bibinfo{pages}{437--440}.
\newblock


\bibitem[Das(2024)]%
        {das2024identity}
\bibfield{author}{\bibinfo{person}{Dipto Das}.}
  \bibinfo{year}{2024}\natexlab{}.
\newblock \emph{\bibinfo{title}{Identity Decolonization amid the Coloniality of
  Computing}}.
\newblock \bibinfo{thesistype}{Ph.\,D. Dissertation}.
  \bibinfo{school}{University of Colorado Boulder}.
\newblock


\bibitem[Das et~al\mbox{.}(2024a)]%
        {das2024reimagining}
\bibfield{author}{\bibinfo{person}{Dipto Das}, \bibinfo{person}{Dhwani Gandhi},
  {and} \bibinfo{person}{Bryan Semaan}.} \bibinfo{year}{2024}\natexlab{a}.
\newblock \showarticletitle{Reimagining Communities through Transnational
  Bengali Decolonial Discourse with YouTube Content Creators}.
\newblock \bibinfo{journal}{\emph{Proceedings of the ACM on Human-Computer
  Interaction}} \bibinfo{number}{CSCW} (\bibinfo{year}{2024}),
  \bibinfo{pages}{1--36}.
\newblock


\bibitem[Das et~al\mbox{.}(2024b)]%
        {das2024colonial}
\bibfield{author}{\bibinfo{person}{Dipto Das}, \bibinfo{person}{Shion Guha},
  \bibinfo{person}{Jed~R Brubaker}, {and} \bibinfo{person}{Bryan Semaan}.}
  \bibinfo{year}{2024}\natexlab{b}.
\newblock \showarticletitle{The``Colonial Impulse" of Natural Language
  Processing: An Audit of Bengali Sentiment Analysis Tools and Their
  Identity-based Biases}. In \bibinfo{booktitle}{\emph{Proceedings of the CHI
  Conference on Human Factors in Computing Systems}}. \bibinfo{pages}{1--18}.
\newblock


\bibitem[Das et~al\mbox{.}(2023)]%
        {das2023toward}
\bibfield{author}{\bibinfo{person}{Dipto Das}, \bibinfo{person}{Shion Guha},
  {and} \bibinfo{person}{Bryan Semaan}.} \bibinfo{year}{2023}\natexlab{}.
\newblock \showarticletitle{Toward cultural bias evaluation datasets: The case
  of Bengali gender, religious, and national identity}. In
  \bibinfo{booktitle}{\emph{Proceedings of the First Workshop on Cross-Cultural
  Considerations in NLP (C3NLP)}}. \bibinfo{pages}{68--83}.
\newblock


\bibitem[Das et~al\mbox{.}(2022)]%
        {das2022understanding}
\bibfield{author}{\bibinfo{person}{Dipto Das}, \bibinfo{person}{AKM~Najmul
  Islam}, \bibinfo{person}{SM~Taiabul Haque}, \bibinfo{person}{Jukka Vuorinen},
  {and} \bibinfo{person}{Syed~Ishtiaque Ahmed}.}
  \bibinfo{year}{2022}\natexlab{}.
\newblock \showarticletitle{Understanding the Strategies and Practices of
  Facebook Microcelebrities for Engaging in Sociopolitical Discourses}. In
  \bibinfo{booktitle}{\emph{Proceedings of the 2022 International Conference on
  Information and Communication Technologies and Development}}.
  \bibinfo{pages}{1--19}.
\newblock


\bibitem[Das et~al\mbox{.}(2021)]%
        {das2021jol}
\bibfield{author}{\bibinfo{person}{Dipto Das}, \bibinfo{person}{Carsten
  {\O}sterlund}, {and} \bibinfo{person}{Bryan Semaan}.}
  \bibinfo{year}{2021}\natexlab{}.
\newblock \showarticletitle{" Jol" or" Pani"?: How Does Governance Shape a
  Platform's Identity?}
\newblock \bibinfo{journal}{\emph{Proceedings of the ACM on Human-Computer
  Interaction}} \bibinfo{volume}{5}, \bibinfo{number}{CSCW2}
  (\bibinfo{year}{2021}), \bibinfo{pages}{1--25}.
\newblock


\bibitem[Das and Semaan(2022)]%
        {das2022collaborative}
\bibfield{author}{\bibinfo{person}{Dipto Das} {and} \bibinfo{person}{Bryan
  Semaan}.} \bibinfo{year}{2022}\natexlab{}.
\newblock \showarticletitle{Collaborative identity decolonization as reclaiming
  narrative agency: Identity work of Bengali communities on Quora}. In
  \bibinfo{booktitle}{\emph{Proceedings of the 2022 CHI Conference on Human
  Factors in Computing Systems}}. \bibinfo{pages}{1--23}.
\newblock


\bibitem[Das~Chomok and Dash~Roni(2023)]%
        {das2023blasphemy}
\bibfield{author}{\bibinfo{person}{Gargi Das~Chomok} {and}
  \bibinfo{person}{Saurov Dash~Roni}.} \bibinfo{year}{2023}\natexlab{}.
\newblock \bibinfo{title}{Blasphemy Laws and Human Rights of Religious
  Minorities in Bangladesh}.
\newblock
  \bibinfo{howpublished}{\url{https://blogs.lse.ac.uk/southasia/2023/11/20/blasphemy-laws-and-human-rights-of-religious-minorities-in-bangladesh/}}.
\newblock
\newblock
\shownote{Last accessed: July 5, 2024}.


\bibitem[Dash et~al\mbox{.}(2022)]%
        {dash2022insights}
\bibfield{author}{\bibinfo{person}{Saloni Dash}, \bibinfo{person}{Rynaa
  Grover}, \bibinfo{person}{Gazal Shekhawat}, \bibinfo{person}{Sukhnidh Kaur},
  \bibinfo{person}{Dibyendu Mishra}, {and} \bibinfo{person}{Joyojeet Pal}.}
  \bibinfo{year}{2022}\natexlab{}.
\newblock \showarticletitle{Insights into incitement: A computational
  perspective on dangerous speech on Twitter in India}. In
  \bibinfo{booktitle}{\emph{Proceedings of the 5th ACM SIGCAS/SIGCHI Conference
  on Computing and Sustainable Societies}}. \bibinfo{pages}{103--121}.
\newblock


\bibitem[Dosono and Semaan(2020)]%
        {dosono2020decolonizing}
\bibfield{author}{\bibinfo{person}{Bryan Dosono} {and} \bibinfo{person}{Bryan
  Semaan}.} \bibinfo{year}{2020}\natexlab{}.
\newblock \showarticletitle{Decolonizing tactics as collective resilience:
  Identity work of AAPI communities on Reddit}.
\newblock \bibinfo{journal}{\emph{Proceedings of the ACM on Human-Computer
  interaction}} \bibinfo{volume}{4}, \bibinfo{number}{CSCW1}
  (\bibinfo{year}{2020}), \bibinfo{pages}{1--20}.
\newblock


\bibitem[Dourish and Mainwaring(2012)]%
        {dourish2012ubicomp}
\bibfield{author}{\bibinfo{person}{Paul Dourish} {and} \bibinfo{person}{Scott~D
  Mainwaring}.} \bibinfo{year}{2012}\natexlab{}.
\newblock \showarticletitle{Ubicomp's colonial impulse}. In
  \bibinfo{booktitle}{\emph{Proceedings of the 2012 ACM conference on
  ubiquitous computing}}. \bibinfo{pages}{133--142}.
\newblock


\bibitem[Erikson(1968)]%
        {erikson1968identity}
\bibfield{author}{\bibinfo{person}{Erik~H Erikson}.}
  \bibinfo{year}{1968}\natexlab{}.
\newblock \bibinfo{booktitle}{\emph{Identity: Youth and crisis}}.
\newblock Number~7. \bibinfo{publisher}{WW Norton \& company}.
\newblock


\bibitem[Eubanks and Sherpa(2018)]%
        {eubanks2018we}
\bibfield{author}{\bibinfo{person}{Charlotte Eubanks} {and}
  \bibinfo{person}{Pasang~Yangjee Sherpa}.} \bibinfo{year}{2018}\natexlab{}.
\newblock \showarticletitle{We Are (Are We?) All Indigenous Here, and Other
  Claims about Space, Place, and Belonging in Asia}.
\newblock \bibinfo{journal}{\emph{Verge: Studies in Global Asias}}
  \bibinfo{volume}{4}, \bibinfo{number}{2} (\bibinfo{year}{2018}),
  \bibinfo{pages}{vi--xiv}.
\newblock


\bibitem[Fan and Zhang(2020)]%
        {fan2020digital}
\bibfield{author}{\bibinfo{person}{Jenny Fan} {and} \bibinfo{person}{Amy~X
  Zhang}.} \bibinfo{year}{2020}\natexlab{}.
\newblock \showarticletitle{Digital juries: A civics-oriented approach to
  platform governance}. In \bibinfo{booktitle}{\emph{Proceedings of the 2020
  CHI conference on human factors in computing systems}}.
  \bibinfo{pages}{1--14}.
\newblock


\bibitem[Fanon(2007)]%
        {fanon2007wretched}
\bibfield{author}{\bibinfo{person}{Frantz Fanon}.}
  \bibinfo{year}{2007}\natexlab{}.
\newblock \bibinfo{booktitle}{\emph{The Wretched of the Earth}}.
\newblock \bibinfo{publisher}{Grove Atlantic}.
\newblock


\bibitem[Fanon(2008)]%
        {fanon2008black}
\bibfield{author}{\bibinfo{person}{Frantz Fanon}.}
  \bibinfo{year}{2008}\natexlab{}.
\newblock \bibinfo{booktitle}{\emph{Black skin, white masks}}.
\newblock \bibinfo{publisher}{Grove press}.
\newblock


\bibitem[Farrell(2011)]%
        {farrell2011fat}
\bibfield{author}{\bibinfo{person}{Amy~Erdman Farrell}.}
  \bibinfo{year}{2011}\natexlab{}.
\newblock \bibinfo{booktitle}{\emph{Fat shame: Stigma and the fat body in
  American culture}}.
\newblock \bibinfo{publisher}{NYU Press}.
\newblock


\bibitem[Fiesler et~al\mbox{.}(2023)]%
        {fiesler2023chilling}
\bibfield{author}{\bibinfo{person}{Casey Fiesler}, \bibinfo{person}{Joshua
  Paup}, {and} \bibinfo{person}{Corian Zacher}.}
  \bibinfo{year}{2023}\natexlab{}.
\newblock \showarticletitle{Chilling Tales: Understanding the Impact of
  Copyright Takedowns on Transformative Content Creators}.
\newblock \bibinfo{journal}{\emph{Proceedings of the ACM on Human-Computer
  Interaction}} \bibinfo{volume}{7}, \bibinfo{number}{CSCW2}
  (\bibinfo{year}{2023}), \bibinfo{pages}{1--21}.
\newblock


\bibitem[Gecas(1982)]%
        {gecas1982self}
\bibfield{author}{\bibinfo{person}{Viktor Gecas}.}
  \bibinfo{year}{1982}\natexlab{}.
\newblock \showarticletitle{The self-concept}.
\newblock \bibinfo{journal}{\emph{Annual review of sociology}}
  \bibinfo{volume}{8}, \bibinfo{number}{1} (\bibinfo{year}{1982}),
  \bibinfo{pages}{1--33}.
\newblock


\bibitem[Geiger(2009)]%
        {geiger2009does}
\bibfield{author}{\bibinfo{person}{R~Stuart Geiger}.}
  \bibinfo{year}{2009}\natexlab{}.
\newblock \showarticletitle{Does Habermas understand the internet? The
  algorithmic construction of the blogo/public sphere}.
\newblock \bibinfo{journal}{\emph{Gnovis. A Journal of Communication, Culture,
  and Technology}} \bibinfo{volume}{10}, \bibinfo{number}{1}
  (\bibinfo{year}{2009}), \bibinfo{pages}{1--29}.
\newblock


\bibitem[Geiger(2016)]%
        {geiger2016bot}
\bibfield{author}{\bibinfo{person}{R~Stuart Geiger}.}
  \bibinfo{year}{2016}\natexlab{}.
\newblock \showarticletitle{Bot-based collective blocklists in Twitter: the
  counterpublic moderation of harassment in a networked public space}.
\newblock \bibinfo{journal}{\emph{Information, Communication \& Society}}
  \bibinfo{volume}{19}, \bibinfo{number}{6} (\bibinfo{year}{2016}),
  \bibinfo{pages}{787--803}.
\newblock


\bibitem[Ghosh and Caliskan(2023)]%
        {ghosh2023person}
\bibfield{author}{\bibinfo{person}{Sourojit Ghosh} {and} \bibinfo{person}{Aylin
  Caliskan}.} \bibinfo{year}{2023}\natexlab{}.
\newblock \showarticletitle{'Person'== Light-skinned, Western Man, and
  Sexualization of Women of Color: Stereotypes in Stable Diffusion}.
\newblock \bibinfo{journal}{\emph{arXiv preprint arXiv:2310.19981}}
  (\bibinfo{year}{2023}).
\newblock


\bibitem[Gilbert(2020)]%
        {gilbert2020run}
\bibfield{author}{\bibinfo{person}{Sarah~A Gilbert}.}
  \bibinfo{year}{2020}\natexlab{}.
\newblock \showarticletitle{" I run the world's largest historical outreach
  project and it's on a cesspool of a website." Moderating a Public Scholarship
  Site on Reddit: A Case Study of r/AskHistorians}.
\newblock \bibinfo{journal}{\emph{Proceedings of the ACM on Human-Computer
  Interaction}} \bibinfo{volume}{4}, \bibinfo{number}{CSCW1}
  (\bibinfo{year}{2020}), \bibinfo{pages}{1--27}.
\newblock


\bibitem[Gillespie(2018)]%
        {gillespie2018custodians}
\bibfield{author}{\bibinfo{person}{Tarleton Gillespie}.}
  \bibinfo{year}{2018}\natexlab{}.
\newblock \bibinfo{booktitle}{\emph{Custodians of the Internet: Platforms,
  content moderation, and the hidden decisions that shape social media}}.
\newblock \bibinfo{publisher}{Yale University Press}.
\newblock


\bibitem[Goffman(1959)]%
        {goffman1959presentation}
\bibfield{author}{\bibinfo{person}{Erving Goffman}.}
  \bibinfo{year}{1959}\natexlab{}.
\newblock \bibinfo{booktitle}{\emph{The presentation of self in everyday
  life/Erving Goffman}}.
\newblock \bibinfo{publisher}{University of Edinburgh}.
\newblock


\bibitem[Group(2018)]%
        {minority2018adivasis}
\bibfield{author}{\bibinfo{person}{Minority~Rights Group}.}
  \bibinfo{year}{2018}\natexlab{}.
\newblock \bibinfo{title}{Adivasis in Bangladesh}.
\newblock
  \bibinfo{howpublished}{\url{https://minorityrights.org/communities/adivasis/}}.
\newblock
\newblock
\shownote{Last accessed: July 21, 2024}.


\bibitem[Guha(2000)]%
        {guha2000unquiet}
\bibfield{author}{\bibinfo{person}{Ramachandra Guha}.}
  \bibinfo{year}{2000}\natexlab{}.
\newblock \bibinfo{booktitle}{\emph{The unquiet woods: ecological change and
  peasant resistance in the Himalaya}}.
\newblock \bibinfo{publisher}{Univ of California Press}.
\newblock


\bibitem[Habermas(2020)]%
        {habermas2020public}
\bibfield{author}{\bibinfo{person}{J{\"u}rgen Habermas}.}
  \bibinfo{year}{2020}\natexlab{}.
\newblock \showarticletitle{The public sphere: An encyclopedia article}.
\newblock In \bibinfo{booktitle}{\emph{Critical theory and society}}.
  \bibinfo{publisher}{Routledge}, \bibinfo{pages}{136--142}.
\newblock


\bibitem[Hampton(2021)]%
        {hampton2021black}
\bibfield{author}{\bibinfo{person}{Lelia~Marie Hampton}.}
  \bibinfo{year}{2021}\natexlab{}.
\newblock \showarticletitle{Black feminist musings on algorithmic oppression}.
\newblock \bibinfo{journal}{\emph{arXiv preprint arXiv:2101.09869}}
  (\bibinfo{year}{2021}).
\newblock


\bibitem[Haque et~al\mbox{.}(2020)]%
        {haque2020privacy}
\bibfield{author}{\bibinfo{person}{SM~Taiabul Haque},
  \bibinfo{person}{MD~Romael Haque}, \bibinfo{person}{Swapnil Nandy},
  \bibinfo{person}{Priyank Chandra}, \bibinfo{person}{Mahdi~Nasrullah
  Al-Ameen}, \bibinfo{person}{Shion Guha}, {and}
  \bibinfo{person}{Syed~Ishtiaque Ahmed}.} \bibinfo{year}{2020}\natexlab{}.
\newblock \showarticletitle{Privacy vulnerabilities in public digital service
  centers in Dhaka, Bangladesh}. In \bibinfo{booktitle}{\emph{Proceedings of
  the 2020 International Conference on Information and Communication
  Technologies and Development}}. \bibinfo{pages}{1--12}.
\newblock


\bibitem[Hill and Chakma(2022)]%
        {hill2022muscular}
\bibfield{author}{\bibinfo{person}{Glen Hill} {and} \bibinfo{person}{Kabita
  Chakma}.} \bibinfo{year}{2022}\natexlab{}.
\newblock \showarticletitle{Muscular nationalism, masculinist militarism: the
  creation of situational motivators and opportunities for violence against the
  Indigenous peoples of the Chittagong Hill Tracts, Bangladesh}.
\newblock \bibinfo{journal}{\emph{International Feminist Journal of Politics}}
  \bibinfo{volume}{24}, \bibinfo{number}{4} (\bibinfo{year}{2022}),
  \bibinfo{pages}{519--543}.
\newblock


\bibitem[Holston(2007)]%
        {holston2007insurgent}
\bibfield{author}{\bibinfo{person}{James Holston}.}
  \bibinfo{year}{2007}\natexlab{}.
\newblock \bibinfo{booktitle}{\emph{Insurgent citizenship: Disjunctions of
  democracy and modernity in Brazil}}.
\newblock \bibinfo{publisher}{Princeton university press}.
\newblock


\bibitem[Hunter(2007)]%
        {hunter2007persistent}
\bibfield{author}{\bibinfo{person}{Margaret Hunter}.}
  \bibinfo{year}{2007}\natexlab{}.
\newblock \showarticletitle{The persistent problem of colorism: Skin tone,
  status, and inequality}.
\newblock \bibinfo{journal}{\emph{Sociology compass}} \bibinfo{volume}{1},
  \bibinfo{number}{1} (\bibinfo{year}{2007}), \bibinfo{pages}{237--254}.
\newblock


\bibitem[Irani et~al\mbox{.}(2010)]%
        {irani2010postcolonial}
\bibfield{author}{\bibinfo{person}{Lilly Irani}, \bibinfo{person}{Janet
  Vertesi}, \bibinfo{person}{Paul Dourish}, \bibinfo{person}{Kavita Philip},
  {and} \bibinfo{person}{Rebecca~E Grinter}.} \bibinfo{year}{2010}\natexlab{}.
\newblock \showarticletitle{Postcolonial computing: a lens on design and
  development}. In \bibinfo{booktitle}{\emph{Proceedings of the SIGCHI
  conference on human factors in computing systems}}.
  \bibinfo{pages}{1311--1320}.
\newblock


\bibitem[(IWGIA)(2021)]%
        {iwgia2021violence}
\bibfield{author}{\bibinfo{person}{International Work Group for
  Indigenous~Affairs (IWGIA)}.} \bibinfo{year}{2021}\natexlab{}.
\newblock \bibinfo{title}{Fact Sheet: Violence against Indigenous women and
  girls in Bangladesh}.
\newblock
  \bibinfo{howpublished}{\url{https://iwgia.org/en/bangladesh/4575-fact-sheet-violence-against-indigenous-women-and-girls-in-bangladesh.html}}.
\newblock
\newblock
\shownote{Last accessed: June 17, 2024}.


\bibitem[Jiang et~al\mbox{.}(2023)]%
        {jiang2023trade}
\bibfield{author}{\bibinfo{person}{Jialun~Aaron Jiang}, \bibinfo{person}{Peipei
  Nie}, \bibinfo{person}{Jed~R Brubaker}, {and} \bibinfo{person}{Casey
  Fiesler}.} \bibinfo{year}{2023}\natexlab{}.
\newblock \showarticletitle{A trade-off-centered framework of content
  moderation}.
\newblock \bibinfo{journal}{\emph{ACM Transactions on Computer-Human
  Interaction}} \bibinfo{volume}{30}, \bibinfo{number}{1}
  (\bibinfo{year}{2023}), \bibinfo{pages}{1--34}.
\newblock


\bibitem[Jiang et~al\mbox{.}(2021)]%
        {jiang2021understanding}
\bibfield{author}{\bibinfo{person}{Jialun~Aaron Jiang},
  \bibinfo{person}{Morgan~Klaus Scheuerman}, \bibinfo{person}{Casey Fiesler},
  {and} \bibinfo{person}{Jed~R Brubaker}.} \bibinfo{year}{2021}\natexlab{}.
\newblock \showarticletitle{Understanding international perceptions of the
  severity of harmful content online}.
\newblock \bibinfo{journal}{\emph{PloS one}} \bibinfo{volume}{16},
  \bibinfo{number}{8} (\bibinfo{year}{2021}), \bibinfo{pages}{e0256762}.
\newblock


\bibitem[Koshy et~al\mbox{.}(2023)]%
        {koshy2023measuring}
\bibfield{author}{\bibinfo{person}{Vinay Koshy}, \bibinfo{person}{Tanvi
  Bajpai}, \bibinfo{person}{Eshwar Chandrasekharan}, \bibinfo{person}{Hari
  Sundaram}, {and} \bibinfo{person}{Karrie Karahalios}.}
  \bibinfo{year}{2023}\natexlab{}.
\newblock \showarticletitle{Measuring User-Moderator Alignment on
  r/ChangeMyView}.
\newblock \bibinfo{journal}{\emph{Proceedings of the ACM on Human-Computer
  Interaction}} \bibinfo{volume}{7}, \bibinfo{number}{CSCW2}
  (\bibinfo{year}{2023}), \bibinfo{pages}{1--36}.
\newblock


\bibitem[Kumar et~al\mbox{.}(2018)]%
        {kumar2018uber}
\bibfield{author}{\bibinfo{person}{Neha Kumar}, \bibinfo{person}{Nassim
  Jafarinaimi}, {and} \bibinfo{person}{Mehrab Bin~Morshed}.}
  \bibinfo{year}{2018}\natexlab{}.
\newblock \showarticletitle{Uber in Bangladesh: The Tangled Web of mobility and
  justice}.
\newblock \bibinfo{journal}{\emph{Proceedings of the ACM on Human-Computer
  Interaction}} \bibinfo{volume}{2}, \bibinfo{number}{CSCW}
  (\bibinfo{year}{2018}), \bibinfo{pages}{1--21}.
\newblock


\bibitem[Lampe et~al\mbox{.}(2006)]%
        {lampe2006face}
\bibfield{author}{\bibinfo{person}{Cliff Lampe}, \bibinfo{person}{Nicole
  Ellison}, {and} \bibinfo{person}{Charles Steinfield}.}
  \bibinfo{year}{2006}\natexlab{}.
\newblock \showarticletitle{A Face (book) in the crowd: Social searching vs.
  social browsing}. In \bibinfo{booktitle}{\emph{Proceedings of the 2006 20th
  anniversary conference on Computer supported cooperative work}}.
  \bibinfo{pages}{167--170}.
\newblock


\bibitem[Lampe et~al\mbox{.}(2014)]%
        {lampe2014crowdsourcing}
\bibfield{author}{\bibinfo{person}{Cliff Lampe}, \bibinfo{person}{Paul Zube},
  \bibinfo{person}{Jusil Lee}, \bibinfo{person}{Chul~Hyun Park}, {and}
  \bibinfo{person}{Erik Johnston}.} \bibinfo{year}{2014}\natexlab{}.
\newblock \showarticletitle{Crowdsourcing civility: A natural experiment
  examining the effects of distributed moderation in online forums}.
\newblock \bibinfo{journal}{\emph{Government Information Quarterly}}
  \bibinfo{volume}{31}, \bibinfo{number}{2} (\bibinfo{year}{2014}),
  \bibinfo{pages}{317--326}.
\newblock


\bibitem[Lenhart et~al\mbox{.}(2024)]%
        {lenhart2024contentr}
\bibfield{author}{\bibinfo{person}{Anna Lenhart}, \bibinfo{person}{Sarah
  Gilbert}, {and} \bibinfo{person}{Katie Shilton}.}
  \bibinfo{year}{2024}\natexlab{}.
\newblock \showarticletitle{CONTENTR: An Experiential Game for Teaching Value
  Tradeoffs in Social Media Governance}. In
  \bibinfo{booktitle}{\emph{Proceedings of the 55th ACM Technical Symposium on
  Computer Science Education V. 1}}. \bibinfo{pages}{722--728}.
\newblock


\bibitem[Leung(2015)]%
        {leung2015validity}
\bibfield{author}{\bibinfo{person}{Lawrence Leung}.}
  \bibinfo{year}{2015}\natexlab{}.
\newblock \showarticletitle{Validity, reliability, and generalizability in
  qualitative research}.
\newblock \bibinfo{journal}{\emph{Journal of family medicine and primary care}}
  \bibinfo{volume}{4}, \bibinfo{number}{3} (\bibinfo{year}{2015}),
  \bibinfo{pages}{324--327}.
\newblock


\bibitem[Liang et~al\mbox{.}(2021)]%
        {liang2021embracing}
\bibfield{author}{\bibinfo{person}{Calvin~A Liang}, \bibinfo{person}{Sean~A
  Munson}, {and} \bibinfo{person}{Julie~A Kientz}.}
  \bibinfo{year}{2021}\natexlab{}.
\newblock \showarticletitle{Embracing four tensions in human-computer
  interaction research with marginalized people}.
\newblock \bibinfo{journal}{\emph{ACM Transactions on Computer-Human
  Interaction (TOCHI)}} \bibinfo{volume}{28}, \bibinfo{number}{2}
  (\bibinfo{year}{2021}), \bibinfo{pages}{1--47}.
\newblock


\bibitem[Loomba(2002)]%
        {loomba2002colonialism}
\bibfield{author}{\bibinfo{person}{Ania Loomba}.}
  \bibinfo{year}{2002}\natexlab{}.
\newblock \bibinfo{booktitle}{\emph{Colonialism/postcolonialism}}.
\newblock \bibinfo{publisher}{Routledge}.
\newblock


\bibitem[Ma et~al\mbox{.}(2023)]%
        {ma2023users}
\bibfield{author}{\bibinfo{person}{Renkai Ma}, \bibinfo{person}{Yue You},
  \bibinfo{person}{Xinning Gui}, {and} \bibinfo{person}{Yubo Kou}.}
  \bibinfo{year}{2023}\natexlab{}.
\newblock \showarticletitle{How Do Users Experience Moderation?: A Systematic
  Literature Review}.
\newblock \bibinfo{journal}{\emph{Proceedings of the ACM on Human-Computer
  Interaction}} \bibinfo{volume}{7}, \bibinfo{number}{CSCW2}
  (\bibinfo{year}{2023}), \bibinfo{pages}{1--30}.
\newblock


\bibitem[McDonald et~al\mbox{.}(2019)]%
        {mcdonald2019reliability}
\bibfield{author}{\bibinfo{person}{Nora McDonald}, \bibinfo{person}{Sarita
  Schoenebeck}, {and} \bibinfo{person}{Andrea Forte}.}
  \bibinfo{year}{2019}\natexlab{}.
\newblock \showarticletitle{Reliability and inter-rater reliability in
  qualitative research: Norms and guidelines for CSCW and HCI practice}.
\newblock \bibinfo{journal}{\emph{Proceedings of the ACM on human-computer
  interaction}} \bibinfo{volume}{3}, \bibinfo{number}{CSCW}
  (\bibinfo{year}{2019}), \bibinfo{pages}{1--23}.
\newblock


\bibitem[Mim et~al\mbox{.}(2022)]%
        {mim2022f}
\bibfield{author}{\bibinfo{person}{Nusrat~Jahan Mim},
  \bibinfo{person}{Dipannita Nandi}, \bibinfo{person}{Sadaf~Sumyia Khan}, {and}
  \bibinfo{person}{Arundhuti Dey}.} \bibinfo{year}{2022}\natexlab{}.
\newblock \showarticletitle{F-commerce and Urban Modernities: The Changing
  Terrain of Housing Design in Bangladesh}. In
  \bibinfo{booktitle}{\emph{Proceedings of the 2022 CHI Conference on Human
  Factors in Computing Systems}}. \bibinfo{pages}{1--20}.
\newblock


\bibitem[Ministry~of Law(1972)]%
        {bangladesh1972constitution}
\bibfield{author}{\bibinfo{person}{Government of~Bangladesh Ministry~of Law}.}
  \bibinfo{year}{1972}\natexlab{}.
\newblock \bibinfo{title}{The Constitution of the People‌'s Republic of
  Bangladesh}.
\newblock
  \bibinfo{howpublished}{\url{http://bdlaws.minlaw.gov.bd/act-367.html}}.
\newblock
\newblock
\shownote{Last accessed: July 5, 2024}.


\bibitem[Molina and Sundar(2022)]%
        {molina2022ai}
\bibfield{author}{\bibinfo{person}{Maria~D Molina} {and}
  \bibinfo{person}{S~Shyam Sundar}.} \bibinfo{year}{2022}\natexlab{}.
\newblock \showarticletitle{When AI moderates online content: effects of human
  collaboration and interactive transparency on user trust}.
\newblock \bibinfo{journal}{\emph{Journal of Computer-Mediated Communication}}
  \bibinfo{volume}{27}, \bibinfo{number}{4} (\bibinfo{year}{2022}),
  \bibinfo{pages}{zmac010}.
\newblock


\bibitem[Narr(2022)]%
        {narr2022coloniality}
\bibfield{author}{\bibinfo{person}{Gregory Narr}.}
  \bibinfo{year}{2022}\natexlab{}.
\newblock \showarticletitle{The coloniality of desire: revealing the desire to
  be seen and blind spots leveraged by data colonialism as AI manipulates the
  unconscious for profitable extraction on dating apps.}
\newblock \bibinfo{journal}{\emph{Revista Fronteiras}} \bibinfo{volume}{24},
  \bibinfo{number}{3} (\bibinfo{year}{2022}).
\newblock


\bibitem[Nova et~al\mbox{.}(2018)]%
        {nova2018silenced}
\bibfield{author}{\bibinfo{person}{Fayika~Farhat Nova},
  \bibinfo{person}{Md~Rashidujjaman Rifat}, \bibinfo{person}{Pratyasha Saha},
  \bibinfo{person}{Syed~Ishtiaque Ahmed}, {and} \bibinfo{person}{Shion Guha}.}
  \bibinfo{year}{2018}\natexlab{}.
\newblock \showarticletitle{Silenced voices: Understanding sexual harassment on
  anonymous social media among Bangladeshi people}. In
  \bibinfo{booktitle}{\emph{Companion of the 2018 ACM Conference on Computer
  Supported Cooperative Work and Social Computing}}. \bibinfo{pages}{209--212}.
\newblock


\bibitem[Nova et~al\mbox{.}(2019)]%
        {nova2019online}
\bibfield{author}{\bibinfo{person}{Fayika~Farhat Nova},
  \bibinfo{person}{MD~Rashidujjaman Rifat}, \bibinfo{person}{Pratyasha Saha},
  \bibinfo{person}{Syed~Ishtiaque Ahmed}, {and} \bibinfo{person}{Shion Guha}.}
  \bibinfo{year}{2019}\natexlab{}.
\newblock \showarticletitle{Online sexual harassment over anonymous social
  media in Bangladesh}. In \bibinfo{booktitle}{\emph{Proceedings of the Tenth
  International Conference on Information and Communication Technologies and
  Development}}. \bibinfo{pages}{1--12}.
\newblock


\bibitem[Papacharissi(2002)]%
        {papacharissi2002virtual}
\bibfield{author}{\bibinfo{person}{Zizi Papacharissi}.}
  \bibinfo{year}{2002}\natexlab{}.
\newblock \showarticletitle{The virtual sphere: The internet as a public
  sphere}.
\newblock \bibinfo{journal}{\emph{New media \& society}} \bibinfo{volume}{4},
  \bibinfo{number}{1} (\bibinfo{year}{2002}), \bibinfo{pages}{9--27}.
\newblock


\bibitem[Ramnath(2012)]%
        {ramnath2012decolonizing}
\bibfield{author}{\bibinfo{person}{Maia Ramnath}.}
  \bibinfo{year}{2012}\natexlab{}.
\newblock \bibinfo{booktitle}{\emph{Decolonizing anarchism: An
  antiauthoritarian history of India's liberation struggle}}.
  Vol.~\bibinfo{volume}{3}.
\newblock \bibinfo{publisher}{AK Press}.
\newblock


\bibitem[Rasul(2007)]%
        {rasul2007political}
\bibfield{author}{\bibinfo{person}{Golam Rasul}.}
  \bibinfo{year}{2007}\natexlab{}.
\newblock \showarticletitle{Political ecology of the degradation of forest
  commons in the Chittagong Hill Tracts of Bangladesh}.
\newblock \bibinfo{journal}{\emph{Environmental Conservation}}
  \bibinfo{volume}{34}, \bibinfo{number}{2} (\bibinfo{year}{2007}),
  \bibinfo{pages}{153--163}.
\newblock


\bibitem[Report(2023)]%
        {tbs2023deaths}
\bibfield{author}{\bibinfo{person}{TBS Report}.}
  \bibinfo{year}{2023}\natexlab{}.
\newblock \bibinfo{title}{Deaths, imprisonments and harassment: The
  controversial history of the Digital Security Act}.
\newblock
  \bibinfo{howpublished}{\url{https://www.tbsnews.net/bangladesh/deaths-imprisonments-and-harassment-controversial-history-digital-security-act-678322}}.
\newblock
\newblock
\shownote{Last accessed: July 5, 2024}.


\bibitem[Rifat et~al\mbox{.}(2024a)]%
        {rifat2024combating}
\bibfield{author}{\bibinfo{person}{Mohammad~Rashidujjaman Rifat},
  \bibinfo{person}{Ashratuz~Zavin Asha}, \bibinfo{person}{Shivesh Jadon},
  \bibinfo{person}{Xinyi Yan}, \bibinfo{person}{Shion Guha}, {and}
  \bibinfo{person}{Syed~Ishtiaque Ahmed}.} \bibinfo{year}{2024}\natexlab{a}.
\newblock \showarticletitle{Combating Islamophobia: Compromise, Community, and
  Harmony in Mitigating Harmful Online Content}.
\newblock \bibinfo{journal}{\emph{ACM Transactions on Social Computing}}
  (\bibinfo{year}{2024}).
\newblock


\bibitem[Rifat et~al\mbox{.}(2024b)]%
        {rifat2024politics}
\bibfield{author}{\bibinfo{person}{Mohammad~Rashidujjaman Rifat},
  \bibinfo{person}{Dipto Das}, \bibinfo{person}{Arpon Podder},
  \bibinfo{person}{Mahiratul Jannat}, \bibinfo{person}{Robert Soden},
  \bibinfo{person}{Bryan Semaan}, {and} \bibinfo{person}{Syed~Ishtiaque
  Ahmed}.} \bibinfo{year}{2024}\natexlab{b}.
\newblock \showarticletitle{The Politics of Fear and the Experience of
  Bangladeshi Religious Minority Communities Using Social Media Platforms}.
\newblock \bibinfo{journal}{\emph{Proceedings of the ACM on human-computer
  interaction}} \bibinfo{number}{CSCW} (\bibinfo{year}{2024}),
  \bibinfo{pages}{1--31}.
\newblock
\newblock
\shownote{in press}.


\bibitem[Said(2014)]%
        {said2014orientalism}
\bibfield{author}{\bibinfo{person}{Edward. Said}.}
  \bibinfo{year}{2014}\natexlab{}.
\newblock \bibinfo{booktitle}{\emph{Orientalism}}.
\newblock \bibinfo{publisher}{Knopf Doubleday Publishing Group}.
\newblock


\bibitem[Scheuerman et~al\mbox{.}(2021)]%
        {scheuerman2021framework}
\bibfield{author}{\bibinfo{person}{Morgan~Klaus Scheuerman},
  \bibinfo{person}{Jialun~Aaron Jiang}, \bibinfo{person}{Casey Fiesler}, {and}
  \bibinfo{person}{Jed~R Brubaker}.} \bibinfo{year}{2021}\natexlab{}.
\newblock \showarticletitle{A framework of severity for harmful content
  online}.
\newblock \bibinfo{journal}{\emph{Proceedings of the ACM on Human-Computer
  Interaction}} \bibinfo{volume}{5}, \bibinfo{number}{CSCW2}
  (\bibinfo{year}{2021}), \bibinfo{pages}{1--33}.
\newblock


\bibitem[Schlesinger et~al\mbox{.}(2017)]%
        {schlesinger2017intersectional}
\bibfield{author}{\bibinfo{person}{Ari Schlesinger}, \bibinfo{person}{W~Keith
  Edwards}, {and} \bibinfo{person}{Rebecca~E Grinter}.}
  \bibinfo{year}{2017}\natexlab{}.
\newblock \showarticletitle{Intersectional HCI: Engaging identity through
  gender, race, and class}. In \bibinfo{booktitle}{\emph{Proceedings of the
  2017 CHI conference on human factors in computing systems}}.
  \bibinfo{pages}{5412--5427}.
\newblock


\bibitem[Schneider et~al\mbox{.}(2021)]%
        {schneider2021modular}
\bibfield{author}{\bibinfo{person}{Nathan Schneider},
  \bibinfo{person}{Primavera De~Filippi}, \bibinfo{person}{Seth Frey},
  \bibinfo{person}{Joshua~Z Tan}, {and} \bibinfo{person}{Amy~X Zhang}.}
  \bibinfo{year}{2021}\natexlab{}.
\newblock \showarticletitle{Modular politics: Toward a governance layer for
  online communities}.
\newblock \bibinfo{journal}{\emph{Proceedings of the ACM on Human-Computer
  Interaction}} \bibinfo{volume}{5}, \bibinfo{number}{CSCW1}
  (\bibinfo{year}{2021}), \bibinfo{pages}{1--26}.
\newblock


\bibitem[Seering et~al\mbox{.}(2022)]%
        {seering2022metaphors}
\bibfield{author}{\bibinfo{person}{Joseph Seering}, \bibinfo{person}{Geoff
  Kaufman}, {and} \bibinfo{person}{Stevie Chancellor}.}
  \bibinfo{year}{2022}\natexlab{}.
\newblock \showarticletitle{Metaphors in moderation}.
\newblock \bibinfo{journal}{\emph{New Media \& Society}} \bibinfo{volume}{24},
  \bibinfo{number}{3} (\bibinfo{year}{2022}), \bibinfo{pages}{621--640}.
\newblock


\bibitem[Semaan et~al\mbox{.}(2015a)]%
        {semaan2015navigating}
\bibfield{author}{\bibinfo{person}{Bryan Semaan}, \bibinfo{person}{Heather
  Faucett}, \bibinfo{person}{Scott Robertson}, \bibinfo{person}{Misa Maruyama},
  {and} \bibinfo{person}{Sara Douglas}.} \bibinfo{year}{2015}\natexlab{a}.
\newblock \showarticletitle{Navigating imagined audiences: Motivations for
  participating in the online public sphere}. In
  \bibinfo{booktitle}{\emph{Proceedings of the 18th ACM Conference on Computer
  Supported Cooperative Work \& Social Computing}}.
  \bibinfo{pages}{1158--1169}.
\newblock


\bibitem[Semaan et~al\mbox{.}(2015b)]%
        {semaan2015designing}
\bibfield{author}{\bibinfo{person}{Bryan Semaan}, \bibinfo{person}{Heather
  Faucett}, \bibinfo{person}{Scott~P Robertson}, \bibinfo{person}{Misa
  Maruyama}, {and} \bibinfo{person}{Sara Douglas}.}
  \bibinfo{year}{2015}\natexlab{b}.
\newblock \showarticletitle{Designing political deliberation environments to
  support interactions in the public sphere}. In
  \bibinfo{booktitle}{\emph{Proceedings of the 33rd Annual ACM Conference on
  Human Factors in Computing Systems}}. \bibinfo{pages}{3167--3176}.
\newblock


\bibitem[Shahid and Vashistha(2023)]%
        {shahid2023decolonizing}
\bibfield{author}{\bibinfo{person}{Farhana Shahid} {and}
  \bibinfo{person}{Aditya Vashistha}.} \bibinfo{year}{2023}\natexlab{}.
\newblock \showarticletitle{Decolonizing Content Moderation: Does Uniform
  Global Community Standard Resemble Utopian Equality or Western Power
  Hegemony?}. In \bibinfo{booktitle}{\emph{Proceedings of the 2023 CHI
  Conference on Human Factors in Computing Systems}}. \bibinfo{pages}{1--18}.
\newblock


\bibitem[Sinha(2017)]%
        {sinha2017colonial}
\bibfield{author}{\bibinfo{person}{Mrinalini Sinha}.}
  \bibinfo{year}{2017}\natexlab{}.
\newblock \bibinfo{booktitle}{\emph{Colonial masculinity: The ‘manly
  Englishman’and the ‘effeminate Bengali’in the late nineteenth
  century}}.
\newblock \bibinfo{publisher}{Manchester University Press}.
\newblock


\bibitem[Spivak(2003)]%
        {spivak2003can}
\bibfield{author}{\bibinfo{person}{Gayatri~Chakravorty Spivak}.}
  \bibinfo{year}{2003}\natexlab{}.
\newblock \showarticletitle{Can the subaltern speak?}
\newblock \bibinfo{journal}{\emph{Die Philosophin}} \bibinfo{volume}{14},
  \bibinfo{number}{27} (\bibinfo{year}{2003}), \bibinfo{pages}{42--58}.
\newblock


\bibitem[Srnicek(2017)]%
        {srnicek2017platform}
\bibfield{author}{\bibinfo{person}{Nick Srnicek}.}
  \bibinfo{year}{2017}\natexlab{}.
\newblock \bibinfo{booktitle}{\emph{Platform capitalism}}.
\newblock \bibinfo{publisher}{John Wiley \& Sons}.
\newblock


\bibitem[Stokke(2017)]%
        {stokke2017politics}
\bibfield{author}{\bibinfo{person}{Kristian Stokke}.}
  \bibinfo{year}{2017}\natexlab{}.
\newblock \showarticletitle{Politics of citizenship: Towards an analytical
  framework}.
\newblock \bibinfo{journal}{\emph{Norsk Geografisk Tidsskrift-Norwegian Journal
  of Geography}} \bibinfo{volume}{71}, \bibinfo{number}{4}
  (\bibinfo{year}{2017}), \bibinfo{pages}{193--207}.
\newblock


\bibitem[Sultana et~al\mbox{.}(2022)]%
        {sultana2022toleration}
\bibfield{author}{\bibinfo{person}{Sharifa Sultana}, \bibinfo{person}{Rokeya
  Akter}, \bibinfo{person}{Zinnat Sultana}, {and}
  \bibinfo{person}{Syed~Ishtiaque Ahmed}.} \bibinfo{year}{2022}\natexlab{}.
\newblock \showarticletitle{Toleration Factors: The Expectations of Decorum,
  Civility, and Certainty on Rural Social Media}. In
  \bibinfo{booktitle}{\emph{Proceedings of the 2022 International Conference on
  Information and Communication Technologies and Development}}.
  \bibinfo{pages}{1--14}.
\newblock


\bibitem[Sultana et~al\mbox{.}(2018)]%
        {sultana2018design}
\bibfield{author}{\bibinfo{person}{Sharifa Sultana},
  \bibinfo{person}{Fran{\c{c}}ois Guimbreti{\`e}re}, \bibinfo{person}{Phoebe
  Sengers}, {and} \bibinfo{person}{Nicola Dell}.}
  \bibinfo{year}{2018}\natexlab{}.
\newblock \showarticletitle{Design within a patriarchal society: Opportunities
  and challenges in designing for rural women in Bangladesh}. In
  \bibinfo{booktitle}{\emph{Proceedings of the 2018 CHI conference on human
  factors in computing systems}}. \bibinfo{pages}{1--13}.
\newblock


\bibitem[Sultana et~al\mbox{.}(2021)]%
        {sultana2021opaque}
\bibfield{author}{\bibinfo{person}{Sharifa Sultana}, \bibinfo{person}{Ilan
  Mandel}, \bibinfo{person}{Shaid Hasan}, \bibinfo{person}{SM~Raihanul Alam},
  \bibinfo{person}{Khandaker~Reaz Mahmud}, \bibinfo{person}{Zinnat Sultana},
  {and} \bibinfo{person}{Syed~Ishtiaque Ahmed}.}
  \bibinfo{year}{2021}\natexlab{}.
\newblock \showarticletitle{Opaque obstacles: The role of stigma, rumor, and
  superstition in limiting women’s access to computing in rural bangladesh}.
  In \bibinfo{booktitle}{\emph{Proceedings of the 4th ACM SIGCAS Conference on
  Computing and Sustainable Societies}}. \bibinfo{pages}{243--260}.
\newblock


\bibitem[Sun and Ni(2022)]%
        {sun2022design}
\bibfield{author}{\bibinfo{person}{Heng Sun} {and} \bibinfo{person}{Wan Ni}.}
  \bibinfo{year}{2022}\natexlab{}.
\newblock \showarticletitle{Design and Application of an AI-Based Text Content
  Moderation System}.
\newblock \bibinfo{journal}{\emph{Scientific Programming}}
  \bibinfo{volume}{2022} (\bibinfo{year}{2022}).
\newblock


\bibitem[Sundar(2016)]%
        {sundar2016burning}
\bibfield{author}{\bibinfo{person}{Nandini Sundar}.}
  \bibinfo{year}{2016}\natexlab{}.
\newblock \bibinfo{booktitle}{\emph{The burning forest: India's war in
  Bastar}}.
\newblock \bibinfo{publisher}{Juggernaut Books}.
\newblock


\bibitem[Tajfel(1974)]%
        {tajfel1974social}
\bibfield{author}{\bibinfo{person}{Henri Tajfel}.}
  \bibinfo{year}{1974}\natexlab{}.
\newblock \showarticletitle{Social identity and intergroup behaviour}.
\newblock \bibinfo{journal}{\emph{Social science information}}
  \bibinfo{volume}{13}, \bibinfo{number}{2} (\bibinfo{year}{1974}),
  \bibinfo{pages}{65--93}.
\newblock


\bibitem[Thach et~al\mbox{.}(2024)]%
        {thach2024trans}
\bibfield{author}{\bibinfo{person}{Hibby Thach}, \bibinfo{person}{Samuel
  Mayworm}, \bibinfo{person}{Michaelanne Thomas}, {and}
  \bibinfo{person}{Oliver~L Haimson}.} \bibinfo{year}{2024}\natexlab{}.
\newblock \showarticletitle{Trans-centered moderation: Trans technology
  creators and centering transness in platform and community governance:
  Trans-centered moderation}. In \bibinfo{booktitle}{\emph{The 2024 ACM
  Conference on Fairness, Accountability, and Transparency}}.
  \bibinfo{pages}{326--336}.
\newblock


\bibitem[Tuck and Yang(2012)]%
        {tuck2012decolonization}
\bibfield{author}{\bibinfo{person}{Eve Tuck} {and} \bibinfo{person}{Wayne
  Yang}.} \bibinfo{year}{2012}\natexlab{}.
\newblock \showarticletitle{Decolonization is not a metaphor. Decolonization:
  Indigeneity}.
\newblock \bibinfo{journal}{\emph{Education \& Society}} \bibinfo{volume}{1},
  \bibinfo{number}{1} (\bibinfo{year}{2012}), \bibinfo{pages}{1--40}.
\newblock


\bibitem[Vaidya et~al\mbox{.}(2021)]%
        {vaidya2021conceptualizing}
\bibfield{author}{\bibinfo{person}{Sahaj Vaidya}, \bibinfo{person}{Jie Cai},
  \bibinfo{person}{Soumyadeep Basu}, \bibinfo{person}{Azadeh Naderi},
  \bibinfo{person}{Donghee~Yvette Wohn}, {and} \bibinfo{person}{Aritra
  Dasgupta}.} \bibinfo{year}{2021}\natexlab{}.
\newblock \showarticletitle{Conceptualizing Visual Analytic Interventions for
  Content Moderation}. In \bibinfo{booktitle}{\emph{2021 IEEE Visualization
  Conference (VIS)}}. IEEE.
\newblock


\bibitem[Woods(2022)]%
        {woods2022bangladesh}
\bibfield{author}{\bibinfo{person}{Oliver Woods}.}
  \bibinfo{year}{2022}\natexlab{}.
\newblock \bibinfo{title}{Bangladesh's Once-In-A-Century Alcohol Reforms}.
\newblock
  \bibinfo{howpublished}{\url{https://asiabrewersnetwork.com/insight/bangladeshs-once-in-a-century-alcohol-reforms}}.
\newblock
\newblock
\shownote{Last accessed: July 8, 2024}.


\bibitem[Xaxa(1999)]%
        {xaxa1999tribes}
\bibfield{author}{\bibinfo{person}{Virginius Xaxa}.}
  \bibinfo{year}{1999}\natexlab{}.
\newblock \showarticletitle{Tribes as indigenous people of India}.
\newblock \bibinfo{journal}{\emph{Economic and political weekly}}
  (\bibinfo{year}{1999}), \bibinfo{pages}{3589--3595}.
\newblock


\bibitem[Zaman(1982)]%
        {zaman1982crisis}
\bibfield{author}{\bibinfo{person}{MQ Zaman}.} \bibinfo{year}{1982}\natexlab{}.
\newblock \showarticletitle{Crisis in Chittagong Hill Tracts: ethnicity and
  integration}.
\newblock \bibinfo{journal}{\emph{Economic and Political Weekly}}
  (\bibinfo{year}{1982}), \bibinfo{pages}{75--80}.
\newblock


\end{thebibliography}

\end{document}